\documentclass[12pt]{iopart}

\expandafter\let\csname equation*\endcsname\relax
\expandafter\let\csname endequation*\endcsname\relax
\usepackage{amsmath}
\usepackage{amssymb}
\usepackage{iopams} 
\usepackage{graphicx}  
\usepackage{subcaption} 
\usepackage{bbm} 
\graphicspath{{pic/}}  
\usepackage{hyperref}

\usepackage{multirow}

\begin{document}

\title[Gaussian orbital perturbation theory for Schwarzschild geodesics]{Gaussian orbital perturbation theory in  Schwarzschild space-time in terms of elliptic functions}

\author{Oleksii Yanchyshen}

\address{ZARM, University of Bremen, Am Fallturm, 28359 Bremen, Germany}
\address{Bogolyubov Institute for Theoretical Physics, 03143 Kyiv, Ukraine}
\ead{oleksii.yanchyshen@zarm.uni-bremen.de}
\author{Claus L\"ammerzahl}

\address{ZARM, University of Bremen, Am Fallturm, 28359 Bremen, Germany}
\ead{claus.laemmerzahl@zarm.uni-bremen.de}
\vspace{10pt}
\begin{indented}
\item[]\today
\end{indented}

\bibliographystyle{alpha}

\begin{abstract}
General relativistic Gauss  equations for osculating elements for bound orbits under the influence of a perturbing force in an underlying Schwarzschild space-time have been derived in terms of Weierstrass elliptic functions. Thereby, the perturbation forces are restricted to act within the orbital plane only. These equations are analytically solved in linear approximation for several different  perturbations such as cosmological constant perturbation, quantum correction to the Schwarzschild metric, and hybrid Schwarzschild/post-Newtonian $2.5$ order self-force for binary systems in an effective one-body framework.
\end{abstract}

%
\vspace{2pc}

%
%
%

%

\section{Introduction}

The detection of gravitational waves with LIGO and VIRGO opened the new era of gravitational wave astronomy
\cite{bambi2022handbook}, which allows us to explore the physics of Black Holes (BH) and Neutron Stars but also to perform high precision tests of General Relativity (GR) and alternative theories. In the near future projects like  LISA \cite{arun2022new} are expected to measure gravitational waves from supermassive BH mergers and also of extreme mass-ratio inspirals (EMRIs). In order to describe theoretically such inspirals we need to develop various perturbation techniques for near horizon phenomena \cite{barack2019black}.    

Our main aim in this work is to set up a GR version of the Gaussian perturbation equations for osculating elements using  analytic solutions of the geodesic equations in a certain space-time as reference. One motivation for that is to set up an efficient description of near BH horizon orbits or for orbits in coalescing binary systems. Then the perturbation might be an extra force due to the deformation of neutron stars or BHs, the radiation reaction force, or forces related to modified gravity. In this work we consider the simplest non-trivial case of bound Schwarzschild geodesics with additional perturbation forces which act only within the orbital plane. For this case we derive perturbation equations and solve these equations for several perturbation forces. 

A full set of the Gaussian perturbation equations in Newtonian gravity usually consists of equations for the following six orbital elements: semi-major axis $a$, eccentricity $e$, inclination $i$, argument of pericenter $\omega$, longitude of the node $\Omega$ and the starting point of the mean anomaly $M$. The condition that the orbital elements do not change the unperturbed form of the formulae for the position as well as for the velocity makes the orbital elements to osculating elements. That means that the perturbation force does not induce any explicit time-dependence in the position and the velocity. In the case when the perturbing force is restricted to be in the orbital plane, the inclination $i$ and longitude of the node $\Omega$ are constant, so that we have four osculating elements only.
In GR the set of osculating elements may consist of, for example, $e$,  semi-latus rectum $p$, the value of relativistic anomaly at pericenter $\chi_0$, $i$, $\Omega$, the initial values $\Phi$ and $T$ of azimuthal angle $\phi$ and coordinate time $t$, see, e.g.,  \cite{warburton2017evolution}. In the special case of a motion not changing the orbital plane, one has $e$, $p$, $\chi_0$, $\Phi$, and $T$, see \cite{PP2007}. 
In our approach we use several equivalent sets of osculating elements: either the invariants of Weierstrass elliptic function $g_2$ and $g_3$, the argument of pericenter $\phi_0$ and GR "mean anomalies" $M_s$ and $M_t$,  or constants of integration $C_1$ and $C_2$ which in non-relativistic limit are related to angular momentum and energy, $\phi_0$, $M_s$ and $M_t$. We use $M_s=\frac{C_1}{r_g^2} (s-s_0)$ and $M_t=\frac{C_1}{C_2 r_g^2} (t-t_0) $ instead of the initial values $s_0$ and $t_0$ of the proper time $s$ and the coordinate time $t$. In the case of forces that do not depend explicitly on the proper and the coordinate times, we can evaluate $M_s$ and $M_t$ by their definitions instead of solving the corresponding equations, which are quite cumbersome. This is in contrast to choosing $t_0$ and $s_0$ as osculating elements, where we cannot omit solving the corresponding equations. Our choice of $M_s$ and $M_t$ over  $s_0$ and $t_0$ has some limitations, as we cannot easily extract effects of the external forces on the dynamics of the coordinate time $t$ itself but only the difference $t-t_0$.

The GR Gaussian perturbation equations are non-linear and coupled, so it is possible to have analytical solutions only perturbatively in linear or higher order approximations, which may be relevant for small forces. In contrast to \cite{PP2007} and similar approaches (\cite{warburton2017evolution}, \cite{osburn2016highly}, \cite{warburton2012evolution}) where the perturbation equations are solved numerically, we obtain analytical expressions for the osculating elements in linear approximation and, in particular, for the secular perturbations. Furthermore, we believe that due to the explicit  connection between $r$ and $\phi$ the use of Weierstrass elliptic functions might be more convenient even for numerical computations compared to the Chandrasekhar $\chi$ parametrisation. It is also very helpful that, using, e.g.,  Wolfram Mathematica, we can evaluate expressions written in terms of Weierstrass elliptic functions with an arbitrary precision. 

As an application of our new technique, we solve the perturbation equations for the force induced by the presence of the cosmological constant in the Schwarzschild--de Sitter metric, for the hybrid Schwarzschild/post-Newtonian $2.5$ order self-force defined in \cite{PP2007} and for quantum gravitational correction to a Schwarzschild BH obtained in  \cite{CK2021}. In addition, we compare pericenter shifts in Schwarzschild--de Sitter space-time obtained using our approach with the analytical solution from \cite{hackmann2008geodesic}. This new scheme can also be applied to space-times with
multipole moments, either for testing space-times solutions for modified gravitational theories or for applying this to satellite orbits in the frame of GR geodesy in analogy to the discussion of 
perturbations of Kepler orbits in the gravitational field of the real Earth \cite{Kaula}.

In comparison with post-Newtonian methods (see reviews \cite{isoyama2020post} or \cite{futamase2007post}) our approach does not have $\frac{r_g}{r}$ as a small parameter, so that in principle it must work closer to a BH horizon than the post-Newtonian methods based on Gaussian perturbation equations. We compare our method with two different post-Newtonian results for the pericenter shift in a Schwarzschild--de Sitter space-time: one obtained as a 
 post-Newtonian asymptotic of an exact solution \cite{hackmann2012observables} and the second obtained from solving a post-Newtonian version of Gaussian equations \cite{kerr2003standard}. From this comparison, we see that our method works better than the second approach and as well as the first one for the situation when the body is close to the BH horizon. All three methods' accuracy exceeds the current observational accuracy for the Solar System. Also, we can use results from post-Newtonian calculations in order to build hybrid schemes (as in \cite{kidder1993coalescing}) from which we can define our perturbing forces (as in \cite{PP2007}).

As in the non-relativistic case when there are two integrals of motion related to time translation invariance and spatial rotation symmetry, there are no secular corrections to the pericenter $r_p$ and apocenter $r_a$ distances  (or equivalently to eccentricity $e$ and semi-latus rectum $p$). Similarly, in the relativistic case if the metric is independent of the coordinate time $t$ and the azimuthal angle $\phi$, then there are two conserved quantities, which lead to the absence of secular corrections to the pericenter $r_p$ and apocenter $r_a$ distances. Due to this, in the cases of quantum corrections and the cosmological constant induced force discussed later, we have secular corrections only to the pericenter shift.

The paper is organised as follows. In section \ref{geodesy::zero} we introduce solutions of the Schwarzschild geodesic equations which play the role of the zeroth order solutions of our perturbation technique. We also define constants of integration which in the next sections will be treated as osculating elements. In section \ref{geodesy::equations}  we  derive the GR Gaussian perturbation equations for bound Schwarzschild geodesics and perturbation forces acting within the orbital plane only. Then in section \ref{solv::lin} we describe the strategy of solving our equations with small perturbing forces. In sections  \ref{Cosmological::sol}, \ref{CK::21::sol} and \ref{PP::07::solve} we solve our perturbation equations for the force induced by the cosmological constant in the Schwarzschild--de Sitter space-time, for  quantum gravitational correction to a Schwarzschild space-time, and for hybrid Schwarzschild/post-Newtonian $2.5$ order self-force. In addition, in section \ref{Cosmological::sol} we compare the secular corrections with the analytically given motion in a Schwarzschild--de Sitter space-time obtained in  \cite{hackmann2008geodesic}. In section \ref{disc::conc} we discuss our results. In \ref{eliptic::func} we give some necessary definitions of elliptic functions, and in  \ref{Hagihara::Sharf} we prove the equivalence of two different exact solutions of Schwarzschild geodesics in terms of Weierstrass elliptic functions.  \ref{table_of_int} gives useful integrals which appear in calculations of sections \ref{Cosmological::sol},  \ref{CK::21::sol} and \ref{PP::07::solve}. Finally, in  \ref{exact::expressions} we present some cumbersome explicit expressions for sections \ref{geodesy::equations}, \ref{Cosmological::sol}, \ref{CK::21::sol} and \ref{PP::07::solve}.

 \section{Geodesic equations in a Schwarzschild space-time and their solutions in terms of Weierstrass elliptic functions } \label{geodesy::zero}
The motion of a test particle in GR is defined by the geodesic equations:
\begin{equation}
	\frac{d^2 x^i}{d s^2} + \Gamma_{kl}^{i} \frac{dx^k}{ds} \frac{dx^l}{ds} =0,
\end{equation}
where $ \Gamma_{kl}^{i}=\frac{1}{2} g^{ij} \left(\partial_k g_{jl} + \partial_l g_{jk} - \partial_j g_{kl}\right)$ are Christoffel symbols, $g_{ij}$ is the space-time metric with signature $(+,-,-,-)$,  $ds = \sqrt{g_{ij} dx^i dx^j}$ is the proper time interval, and the indices run from 0 to 3. The Schwarzschild metric is given by
\begin{equation}
	g_{00} = 1-\frac{r_g}{r}\, , \quad 
	g_{11} = -  \frac{1}{1-\frac{r_g}{r}}\, , \quad
	g_{22}= - r^2 \, , \quad
	g_{33} = - r^2 \sin^2 \theta \, . 
\end{equation}
The velocity of light $c=1$, and $r_g$ is the Schwarzschild radius. 

In this paper we restrict to forces which act within the orbital plane, so that for convenience we can fix $\theta = \frac{\pi}{2}$. The corresponding set of geodesic equations then is
\begin{align}
	0 & = \ddot{t}  +\frac{   r_g }{r(r- r_g)} \dot{t} \dot{r} \label{eqv::t_s}\\	
 0 & = \ddot{r} +\frac{ \left(r-r_g\right) r_g}{2 r^3} \dot{t}^2- \frac{ r_g}{2 r( r- r_g)}\dot{r}^2-\left(r-r_g\right)  \dot{\phi}^2 \label{eqv::r_s}\\
0 & = \ddot{ \phi}+\frac{2  }{r} \dot{ \phi}\dot{ r} \, ,\label{eqv::phi_s}
\end{align}
 where an overdot denotes a derivative with respect to the proper time $s$.

The first integration of (\ref{eqv::phi_s}) and (\ref{eqv::t_s})  gives 
\begin{align}
\dot{\phi} & =  C_1 \frac{1}{r^2} \label{eqv::phi_s_c1}\\
\dot{t} & = C_2 \frac{r}{r-r_g} \, ,\label{eqv::t_s_c2}
\end{align}
 where $C_1$ and $C_2$ are constants of motion which we are going to treat as osculating elements in the next section. Instead of solving equation  (\ref{eqv::r_s}) for $r(s)$ it is more convenient  to find $r(\phi)$.
For the connection between $r $ and $\phi $ we have an equation as an algebraic curve of genus one, which can be parametrised in terms of the Weierstrass elliptic function~$\wp$~\cite{Hagihara,hackmann2010geodesic}
	\begin{equation}
		r = \frac{ r_g }{ \wp(\frac12 (\phi+\phi_0) + \omega_3) + \frac{1}{3}},\label{hagihara::sol:r::phi}
	\end{equation}
where $\phi_0$ is a further constant of integration (which we also will treat as an osculating element in the next section) and where $\omega_3$ is an imaginary half-period of $\wp$ which is necessary in order for $\wp $ to be a real-valued function. The invariants of $\wp $, $g_2$ and $g_3$, play the role of geometrical characteristics of the trajectory and are related to $C_1$ and $C_2$ according to
\begin{align}
g_2 &= 4  \left( \frac{1}{3}-\frac{ r_g^2}{C_1^2}\right),\label{conection::g2_c1}\\
g_3 &= \frac{4}{3} \left( \frac{2}{9}-\frac{ \left(3 C_2^2-2\right) r_g^2}{ C_1^2}\right)\label{conection::g3_c2}.
\end{align}
For more information about Weierstrass functions see \ref{eliptic::func} or consult classic textbooks like \cite{WW}.
 This is not the only  way to write a solution of the equations of motion (\ref{eqv::t_s})-(\ref{eqv::phi_s}) in terms of the Weierstrass elliptic function, see \cite{Scharf2011},  for example. However, we have shown in \ref{Hagihara::Sharf} that these two approaches are equivalent.

 It is convenient to introduce a “semi-latus rectum” $p$ and an  “eccentricity” $e$  from the definitions of the pericenter $r_p$ and apocenter $r_a$ distances 
 \begin{align}
r_p & = \frac{p}{1+e}\\
r_a & = \frac{p}{1-e} \, ,
\end{align}
which are related to 
the roots $e_2$ and $e_3$ of the Weierstrass $\wp$-function
\begin{align}
e_2 & = \frac{r_g}{r_p}-\frac{1}{3} \\
e_3 & = \frac{r_g}{r_a}-\frac{1}{3} \, , 
\end{align}
that is, 
\begin{align}
e_2 & = \frac{e+1}{p}-\frac{1}{3},\\
e_3 & = \frac{ 1-e}{p}-\frac{1}{3} \, .
\end{align}

From (\ref{eqv::phi_s_c1}) and (\ref{eqv::t_s_c2}) we can define a proper time mean anomaly and a coordinate time mean anomaly as
\begin{equation}
M_s=\frac{C_1}{r_g^2} (s-s_0)  = \int \frac{d\phi}{ \left(\frac{1}{3}+\wp(v)\right)^2 } ,\label{eqv::s_intdph}
\end{equation}
and 
\begin{equation}
M_t=\frac{C_1}{C_2 r_g^2} (t-t_0)   = \int \frac{d\phi}{ \left(\frac{1}{3}+\wp(v)\right)^2 \left(\frac{2}{3}-\wp(v)\right)},\label{eqv::t_intdph}
\end{equation}
where $s_0$ and $t_0$ are constants of integration and where $v= \frac{\phi+\phi_0}{2}+\omega_3$ is defined as the true anomaly. Integration of   (\ref{eqv::s_intdph}) yields
\begin{equation}
M_s = -\frac{2}{\wp'^2(y)} \left( \bar{\xi}_1(\phi,- \tfrac{1}{3}) - \frac{\wp''(y)}{\wp'(y)} \bar{\xi}_2(\phi, - \tfrac13) \right) \, ,\label{eqv::s_intdph2}
\end{equation}
where we denoted $y= \wp^{-1}(- \tfrac13)$ and $\bar{\xi}_j(\phi,\rho)= \xi_j(\phi,\rho)-\xi_j(0,\rho)$ and 
\begin{align} 
\xi_1(\phi,\rho) & = \zeta \left(v-x\right)+\zeta \left(x+v \right)+2 \wp(x)  v \\ 
\xi_2(\phi,\rho) & = 2\zeta (x) v+\log \left(\frac{\sigma \left(v-x\right)}{\sigma \left(x+v \right)}\right) \, ,
\end{align}   
where here and in the following $x= \wp^{-1}(\rho)$, $\zeta$ and $\sigma$ are the Weierstrass zeta and sigma functions. A prime denotes a derivative with respect to the argument. Analogously, for~(\ref{eqv::t_intdph}) we obtain  
\begin{equation}
M_t= M_s +  \frac{2}{\wp'(y)} \bar{\xi}_2(\phi,- \tfrac13)- \frac{2}{\wp'(z)} \bar{\xi}_2(\phi, \tfrac23) \, ,
\end{equation}
where $z= \wp^{-1}(\tfrac23)$. We notice that, while $\xi_j$ are complex-valued functions, the differences~$\bar{\xi}_j$ are  real-valued.
	
\section{Perturbation equations in GR}
\label{geodesy::equations}

The motion of a test particle in GR exposed to an additional force $f^i$ is defined by the equations of
motion
\begin{equation}
		\frac{d^2 x^i}{d s^2} + \Gamma_{kl}^{i} \frac{dx^k}{ds} \frac{dx^l}{ds} =f^i,
\end{equation}
	where  the orthogonality condition
\begin{equation}
		f^i \dot{x}_i =0, \label{eqv::ortogonality::f}
\end{equation}
is satisfied automatically for arbitrary forces 
 provided that the normalisation condition~$\dot{x}_i \dot{x}^i=1$ holds true.
For the Schwarzschild metric together with the perturbation force the equations of motion are
\begin{align}
\ddot{t} + \frac{   r_g }{r(r- r_g)} \dot{t} \dot{r} &= f^t \label{eqv::t_s_f} \\
\ddot{r} +\frac{ \left(r-r_g\right) r_g}{2 r^3} \dot{t}^2- \frac{ r_g}{2 r( r- r_g)}\dot{r}^2-\left(r-r_g\right)  \dot{\phi}^2 &= f^r \label{eqv::r_s_fr}\\
\ddot{ \phi}+\frac{2  }{r} \dot{ \phi}\dot{ r} &= f^\phi \, .\label{eqv::phi_s_f}
\end{align}
 
In order to write perturbation equations for the osculating elements $C_1$, $C_2$ (or~$g_2$,~$g_3$), $v$ (or $\phi_0$) 
we use the technique inspired by the method of variation of constants which is widely used in Newtonian celestial mechanics, see, e.g.,  \cite{Klioner}. First, we postulate that the following relations  (which satisfy the geodesic equations (\ref{eqv::t_s})-(\ref{eqv::phi_s})) hold even in the case of the presence of a perturbation force  
\begin{align}
	\dot{\phi} &= C_1 \frac{1}{r^2},\label{eqv::dot_phi} \\
	\dot{t} &= C_2 \frac{r}{r-r_g},\label{eqv::dot_t}\\
	r &= \frac{ r_g }{ \wp(v) + \frac{1}{3}},\label{eqv::r}\\
	\dot{r} &= -\frac{C_1}{2 r_g } \wp'\left(v\right),\label{eqv::dot_r}
\end{align}
where now the orbital elements are functions of $s$ for which we will obtain equations. The orbital elements fulfilling (\ref{eqv::dot_phi})-(\ref{eqv::dot_r}) are called osculating elements. 

It is necessary to choose an independent set of osculating elements. Such a set can be  $C_1 (s)$, $C_2 (s)$ (or equivalently $g_2(s)$, $g_3(s) $) and  the true anomaly
\begin{align}
	 v(s) = \frac{\phi(s)+\phi_0(s)}{2} +\omega_3(s)  
\end{align}
 (or equivalently  the argument of pericenter  $\phi_0(s)$). 
In order to obtain equations for these osculating elements we take the derivative  of (\ref{eqv::dot_phi}) and (\ref{eqv::dot_t}) with respect to $s$  and eliminate all second order derivatives using (\ref{eqv::phi_s_f})  and (\ref{eqv::t_s_f})
\begin{align}
	\dot{C_1} (s) &= r^2 f^\phi, \label{eqv::dot_c1}\\
	\dot{C_2} (s) &= \left(1- \frac{r_g}{r}\right) f^t, \label{eqv::dot_c2}
\end{align}
or, using  orthogonality condition (\ref{eqv::ortogonality::f}), we  rewrite the latter in terms of $f^\phi$ and $f^r$
	\begin{equation}
		\dot{C_2}(s) = \left(1-\frac{r_g}{r}\right)\frac{ C_1 }{C_2} f^\phi-\frac{ \wp'(v) C_1}{2 r_g C_2} f^r .\label{eqv::dot_c22}
	\end{equation}
 Using connection between $g_2$, $g_3$ and $C_1$, $C_2$ -- 
 (\ref{conection::g2_c1}), (\ref{conection::g3_c2})  we rewrite (\ref{eqv::dot_c1}) and (\ref{eqv::dot_c22}) as
\begin{align}
\dot{g_2}(s) &= A^{g_2}_{\phi} f^\phi, \label{eq::g2s::a}\\
\dot{g_3}(s) &=  A^{g_3}_{r} f^r  + A^{g_3}_{\phi} f^{\phi} \label{eq::g3s::a},
\end{align}
where we denoted
\begin{align}
A^{g_2}_{\phi}(s) &= \frac{ r^2 }{r_g} \left(\frac{4}{3}-g_2\right)^{3/2}, \\
A^{g_3}_r (s)&=  2  \wp'(v)  \left(\frac{4}{3}-g_2\right)^{1/2},\\
A^{g_3}_{\phi} (s)&= -\frac{  \left(r^3 \left(27   g_3-8  \right)+108 r_g^2(r-r_g)\right)}{27 r_g r} \left(\frac{4}{3}-g_2\right)^{1/2} .
\end{align}

In order to obtain an equation for the true anomaly $v$ we take a derivative of~$r(\phi)$~(\ref{eqv::r})  with respect to $s$, and replace  the derivatives $\dot{r}$, $\dot{g}_2$, $\dot{g}_3$, $\frac{\partial \wp}{\partial g_2}$  and  $\frac{\partial \wp}{\partial g_3}$ using~(\ref{eqv::dot_r}), (\ref{eq::g2s::a}), (\ref{eq::g3s::a}) and also  (\ref{eq::part::wp::g2}), (\ref{eq::part::wp::g3}) from \ref{eliptic::func}, and finally get after some simplification (involving some properties of $\wp$ and $\wp'$)  
	\begin{equation}
		\dot{v}(s) = \frac{C_1 }{ 2 r^2} +  A^{v}_r  f^r + A^{v}_{\phi}  f^\phi, \label{eqv::dot_v}
	\end{equation}
	where we also denoted
	\begin{align}
		A_r^v(s) &=   \left(\frac{4}{3}-g_2\right)^{1/2} \frac{  \left(-6 g_2 \left(\zeta (v) \wp'(v)+2 \wp (v)^2\right)+2 g_2^2+9 g_3 \left(v\wp' (v)+2 \wp (v)\right)\right)}{  \left(g_2^3-27 g_3^2\right)}, \\
		A_{\phi}^v(s) &=\frac{3  r_g  }{4 \left(g_2^3-27 g_3^2\right) (3 \wp (v)+1)^2}  \left(\frac{4}{3}-g_2\right)^{1/2}\times \nonumber   \\& \quad \times \Big(  -18 g_3 \left(3 g_3 v-12 v  \wp (v)^3+4 v  \wp (v)-4 \zeta (v)-6 \wp (v) \wp (v)'\right) v\nonumber\\& \qquad\;\;\; +3 g_2^3 -2 g_2^2 \left(2 v+3 \wp (v)'\right) \nonumber\\
  & \qquad\;\;\; -6 g_2 \left(\zeta (v) \left(3 g_3+24 \wp (v)^3-8 \wp (v)\right)+4 \left(3 \wp (v)^2-1\right) \wp (v)'\right) \Big).
	\end{align}
In the case when the force does not depend on the proper time explicitly instead of proper time parametrisation it is convenient to use  $\phi$ as a parameter of motion, so  we   rewrite our perturbations equations (\ref{eqv::dot_c1}) and  (\ref{eqv::dot_c22}) in terms of $\phi$ as
\begin{align}
	C_1'(\phi) &= \frac{r^4}{C_1} f^\phi  ,   \label{eqv::C1_ph}\\
	C_2'(\phi) &=  \left(1-\frac{r_g}{r}\right)\frac{r^2}{C_2}f^\phi -\frac{ r^2 \wp'(v) }{2 r_g C_2} f^r , \label{eqv::C2_ph}
\end{align}
or, analogously, (\ref{eq::g2s::a}) and 
 (\ref{eq::g3s::a}) as
\begin{align}
	g_2'(\phi) = &\frac{ r^4 }{r_g} \left(\frac{4}{3}-g_2\right)^{2} f^\phi, \label{eqv::g2_ph} \\
	g_3'(\phi) = & \frac{r^2 \wp'(v)  }{  r_g} \left(\frac{4}{3}-g_2\right)  f^r \nonumber \\ &-\frac{r    \left(r^3 \left(27  g_3-8\right)+108 r_g^2(r-r_g)\right)}{54 r_g^2 } \left(\frac{4}{3}-g_2\right) f^{\phi} \label{eqv::g3_ph},
\end{align}
and (\ref{eqv::dot_v}) as
\begin{equation}
v'(\phi) = \frac{1 }{ 2} +  B_r^vf^r +B_{\phi}^v f^\phi \label{eqv::v_ph},
\end{equation}
where prime denotes a derivative with respect to $\phi$, and 
\begin{equation}
B_r^v(\phi) = A^v_{r} \frac{r^2}{C_1}, \qquad
B_{\phi}^v(\phi) =A^v_{\phi} \frac{r^2}{C_1}.\label{eqv::brv::def}
\end{equation}

A full set of perturbation equations for osculating parameters must also contain equations for $M_s$ 
and $M_t$. 
But due to their cumbersomeness and the fact that in the case when the force does not depend explicitly on the proper and coordinate times we can find $M_s$ and $M_t$ directly via their definitions~(\ref{eqv::s_intdph}) and~(\ref{eqv::t_intdph}), we will omit writing these equations.

The perturbation equations for osculating elements derived above  (\ref{eq::g2s::a}), (\ref{eq::g3s::a}) and (\ref{eqv::dot_v}) are equivalent to equations obtained in \cite{PP2007} 
\begin{equation}
\frac{\partial z^\alpha}{\partial I_A} \dot{I}_A =0, \qquad
	\frac{\partial \dot{z}^\alpha}{\partial I_A} \dot{I}_A =f^\alpha, \label{eq::PP:main}
\end{equation}
where $z = (t,r,\phi)$ and $I_A$ are five osculating elements.  We decided to write our equations in a manner closer to the standard approach in classical  celestial mechanics (see \cite{Klioner})
because, instead of using the Chandrasekhar $\chi$ parametrisation, we used the more direct connection between $r$ and $\phi$ (\ref{hagihara::sol:r::phi}) which allowed us to omit solving a system of linear algebraic equations  for $\dot{I}_A$ (\ref{eq::PP:main}).  It is easy to show that solutions $\dot{I}_A$ of (\ref{eq::PP:main}) are equivalent to our system of equations (\ref{eq::g2s::a}), (\ref{eq::g3s::a}) and (\ref{eqv::dot_v}).

\section{Solving strategy}\label{solv::lin}
In general, for the forces which do not depend on proper time explicitly we can symbolically write the perturbation equations (\ref{eq::g2s::a}), (\ref{eq::g3s::a}) and (\ref{eqv::dot_v}) as 
\begin{align}
	\dot{g}_2(s) &= F^{g_2}(g_2(s),g_3(s),v(s)),\label{eq::g2::s::F} \\
    \dot{g}_3(s) &= F^{g_3}(g_2(s),g_3(s),v(s)), \label{eq::g3::s::F}\\
    \dot{v}(s) &= F^v(g_2(s),g_3(s),v(s)), \label{eq::v::s::F}
\end{align}
where the $F^{j}(g_2,g_3,v)$ are the RHSs of (\ref{eq::g2s::a}), (\ref{eq::g3s::a}) and (\ref{eqv::dot_v}). This is a set of nonlinear first order differential equations which cannot be solved analytically for arbitrary forces. In the case when the forces are small we can  solve equations  (\ref{eq::g2::s::F})-(\ref{eq::v::s::F})  in linear approximation
\begin{align}
	\dot{g}_2(s) &=   F^{g_2}(g_{2},g_{3},v),\label{eq::g2::s::F::lin} \\
    \dot{g}_3(s) &=  F^{g_3}(g_{2},g_{3},v), \label{eq::g3::s::F::lin}\\
    \dot{v}(s) &=   F^v(g_{2},g_{3},v), \label{eq::v::s::F::lin}
\end{align}
where $g_{2}$, $g_{3}$, and also $ \phi_0$ and $\omega_3$ in the true anomaly $v$ on the RHSs are now constants. Thus, the RHSs depend only on $\phi(s)$ which allows us to  reparametrise them in terms of~$\phi$ 
\begin{align}
	g'_2(\phi) &=  \bar{F}^{g_2}(g_{2},g_{3},v), \\
    g'_3(\phi) &= \bar{F}^{g_3}(g_{2},g_{3},v), \\
    v'(\phi) &=  \bar{F}^v(g_{2},g_{3},v), 
\end{align}
so that it is in principle possible to integrate these equations.

\subsection{Secular corrections}	
 
In order to obtain secular perturbations, we expand the RHSs of (\ref{eq::g2::s::F::lin})-(\ref{eq::v::s::F::lin})  in a proper time Fourier series 
\begin{align}
	\dot{g}_2(s) &= F^{g_2}_0 + \sum_{k\ne 0}^{\infty}  F_{k}^{g_2} e^{ i k M_{s} } , \\
    \dot{g}_3(s) &= F^{g_3}_0+ \sum_{k\ne 0}^{\infty}  F_{k}^{g_3} e^{ i k M_{s} } , \\
    \dot{v}(s) &= F^v_0+  \sum_{k\ne 0}^{\infty}  F_{k}^{v} e^{ i k M_{s} }, 
\end{align}
where $k =\frac{ 2\pi n }{M_s(4 \omega_1) } $ and $n \in N $. $M_s(4 \omega_1)$ is given in \ref{exact::expressions::MsMt}.  We can integrate these equations neglecting all oscillatory terms as 
\begin{align}
	g_2(s) &= g_{2} +  (s-s_0) F^{g_2}_0 , \\
    g_3(s) &= g_{3}+  (s-s_0) F^{g_3}_0 , \\
    v(s) &= v_0 +   (s-s_0) F^v_0   , 
\end{align}
where the zeroth harmonics are given by 
\begin{align}
	F_{0}^{g_2} &= \frac{1}{M_s(4\omega_1)}\int_{P_s} F^{g_2}(s)  d M_{s}, \\
    F_{0}^{g_3} &= \frac{1}{M_s(4\omega_1)}\int_{P_s}   F^{g_3}(s)  d M_{s}, \\
    F_{0}^{v} &= \frac{1}{M_s(4\omega_1)}\int_{P_s}   F^{v}(s)  d M_{s},
\end{align}
where $\int_{P_s}$ is an integral over a period $P_s=M_s(4\omega_1)$.

The same we can do for the coordinate time. We  
first define the coordinate time linearised perturbation equations according to 
\begin{align}
	\dot{g}_2(t) &= G^{g_2}(g_{2},g_{3},v), \\
    \dot{g}_3(t) &= G^{g_3}(g_{2},g_{3},v), \\
    \dot{v}(t) &= G^v(g_{2},g_{3},v),
\end{align}
and then write the coordinate time Fourier series
\begin{align}
	\dot{g}_2(t) &= G^{g_2}_0 + \sum_{k\ne 0}^{\infty}  G_{k}^{g_2} e^{ i k M_{t} } , \\
    \dot{g}_3(t) &= G^{g_3}_0+ \sum_{k\ne 0}^{\infty}  G_{k}^{g_3} e^{ i k M_{t} } , \\
    \dot{v}(t) &= G^v_0+  \sum_{k\ne 0}^{\infty}  G_{k}^{v} e^{ i k M_{t} }, 
\end{align}
where the zeroth harmonics are given by
\begin{align}
	G_{0}^{g_2} &= \frac{1}{M_t(4\omega_1)}\int_{P_t}  G^{g_2}(t)   d M_{t}, \\
    G_{0}^{g_3} &= \frac{1}{M_t(4\omega_1)}\int_{P_t}  G^{g_3}(t)   d M_{t}, \\
    G_{0}^{v} &= \frac{1}{M_t(4\omega_1)}\int_{P_t}  G^{v}(t)   d M_{t},
\end{align}
and $\int_{P_t}$ is an integral over a period $P_t=M_t(4\omega_1)$ which is also given in \ref{exact::expressions::MsMt}.
Thus, we have the secular perturbations
\begin{align}
	g_2(t) &= g_{2} +  (t-t_0) G^{g_2}_0 , \\
    g_3(t) &= g_{3}+  (t-t_0) G^{g_3}_0 , \\
    v(t) &= v_0 +  (t-t_0) G^v_0.
\end{align}
In addition, there is a connection between the proper and coordinate times secular perturbations 
\begin{equation}
G^{j}_0 = \frac{M_s(4\omega_1)}{M_t(4\omega_1) C_2} F^{j}_0.\label{eq::conect:GF}
\end{equation}

Using the same approach we can analogously write the secular perturbations for $C_1(s)$ and $C_2(s)$ instead of $g_2(s)$ and $g_3(s)$. We wrote our general perturbation scheme for $v(s)$  but for calculations with specific forces it is more convenient to use $\phi_0(s)$ or $\bar{\phi}_0(s)= \phi_0(s) +2 \omega_3(s)$ instead. When  $F_0^{C_1}$ and $F_0^{C_2}$ are vanishing there are no secular perturbations of $e$, $p$, $r_a$, $r_p$.  Then the only non-zero secular correction $F_{0}^{\bar{\phi}_0} $ gives an additional pericenter shift per  revolution
\begin{equation}
	\Delta =\frac{M_s(4\omega_1)}{2} \frac{r_g^2}{C_1} ReF^{\bar{\phi}_0}_0 .\label{eq::peri::one}
\end{equation}

\section{Perturbation force from a Schwarzschild--de Sitter metric}	\label{Cosmological::sol}

Though there is an analytical solution for the geodesic equations in the Schwarzschild--de Sitter space-time \cite{Hackmann:2008zza,hackmann2008geodesic} we treat here the influence of the cosmological constant as perturbation and apply our perturbation scheme in order to discuss the performance of the new ansatz. 

The non-vanishing metric elements for the Schwarzschild--de Sitter space-time are 
\begin{equation}
g_{00} = 1-\frac{r_g}{r} - \Lambda    \frac{r^2}{3} ,\quad
g_{11} = - \frac{1}{1-\frac{r_g}{r}-   \Lambda \frac{r^2}{3}}, \quad 
g_{22}=-r^2, \quad
g_{33}=-r^2 \sin^2 \theta,
\end{equation}
where $\Lambda$ is the cosmological constant. Performing a series expansion for small $ \Lambda r_g^2$ of the geodesic equations we obtain the corresponding perturbation force for an underlying Schwarzschild metric 
\begin{align}
f^\phi &=  0  ,\\
 f^t &=  \frac{  \Lambda    r  }{3}  \dot{t} \dot{r} 
 \frac{ (2-3 \frac{r_g}{r})}{  (1-\frac{r_g}{r})^2}  \label{eq::ft::der::lam},
\end{align}
where we do not show the explicit expression for $f^r$ which is quite cumbersome. Using the orthogonality condition $f^i \dot{x}_i=0$, we can express $f^r$ through     $f^t$ and avoid using it completely.
In the following subsections we integrate the linearised  perturbation equations for $C_1(\phi)$, $C_2(\phi)$ and $v(\phi)$  (\ref{eqv::C1_ph}, \ref{eqv::C2_ph}, \ref{eqv::v_ph}) with this force. 

\subsection{Solutions for the osculating elements $C_1$ and $C_2$}
 As postulated in section \ref{geodesy::equations}, we use $\dot{t}$ and $\dot{r}$  as in the homogenous case  (\ref{eqv::dot_t}) and (\ref{eqv::dot_r}). Insertion into $f^t$  (\ref{eq::ft::der::lam})  yields
\begin{equation}
 f^t =  -\frac{\Lambda     C_1 C_2  r }{ 6 r_g } \frac{ \wp'(v)  (2-3\frac{r_g}{r})}{  \left(1-\frac{r_g}{r}\right)^3 } .
\end{equation}
 Equations for $C_1(\phi)$ and $C_2(\phi)$ --- 
 (\ref{eqv::C1_ph}) and (\ref{eqv::C2_ph}) with this force are
\begin{align}
C_1'(\phi) &= 0 ,\\
C'_2 (\phi) &= \left(1-\frac{r_g}{r}\right) \frac{r^2}{C_1} f^t ,
\end{align}
from which we can see that $C_1(\phi) = const $, so we have only one equation 
\begin{equation}
 C_2' (\phi) = -  \frac{\Lambda  r^3 C_2  }{6 r_g} \frac{   (  2 -3 \frac{r_g}{r} )}{    \left(1-\frac{ r_g }{r}\right)^2 }  \wp'(v) . \label{eqv::c2_cosmol}
\end{equation}
We solve (\ref{eqv::c2_cosmol}) in linear approximation by $\Lambda r_g^2$
	\begin{equation}
		C_2(\phi) = C_{2} + \Lambda  r_g^2 \frac{ C_2  }{6 } a(\phi)  ,
	\end{equation}
 where here and throughout the text  $C_2$ and other osculating elements without the argument are constants.
Solving then the equation for $a(\phi)$  
\begin{equation}
a' (\phi) =  -  \frac{r^3}{r_g^3} \frac{(2 - 3 \frac{r_g}{r})}{    \left(1-\frac{r_g}{r}\right)^2} \wp'(v)  ,
\end{equation}
 we have 
\begin{equation}
 a(\phi) =   2 \left(   \alpha(\phi)  - \alpha(0) \right), 
\end{equation}
where 
\begin{equation}
\alpha(\phi) =  \frac{3}{(1+3\wp(v))^2}+ \frac{1}{1+3 \wp(v)}+\frac{1}{2-3 \wp(v)} .
\end{equation}

\subsection{Solution for argument of pericenter }
Next we integrate the equation for $v$ (\ref{eqv::v_ph}) writing it in terms of $f^t$:
\begin{equation}
v'(\phi) = \frac{1 }{ 2} - \left(1-\frac{r_g}{r}\right) \frac{C_2}{C_1} \frac{2 r_g}{ \wp'(v)}B_r^v  f^t  .
\end{equation}
We rewrite this as an equation for  $\bar{\phi}_0(\phi) =\phi_0(\phi)+2\omega_3 $  instead of $v(\phi)$
\begin{eqnarray}
	\bar{\phi}_0'(\phi)= & \Lambda r_g^2  \frac{3 \left(18 g_2+27 g_3-32\right)     }{g_2^3-27 g_3^2 }  \times  \label{eq::ph0bar::cosmol} \\& \times \frac{  (1-3 \wp)  \left(6 g_2 \left(\zeta  \wp '+2 \wp ^2\right)-9 g_3 \left(v \wp '+2 \wp \right)-2 g_2^2\right)}{  (2-3 \wp )^2 (1+3 \wp )^3},\nonumber 
\end{eqnarray}
as it is more convenient to solve. Here we omitted the arguments of $\wp(v)$, $\wp'(v)$ and~$\zeta(v)$ for simplicity. We integrate equation (\ref{eq::ph0bar::cosmol}) in linear approximation, using formulas~(\ref{eq::int_wpP_v_frac2_dv})-(\ref{eq::int_fracn_dv}) from  \ref{table_of_int} 
as
\begin{equation}
\bar{\phi}_0(\phi) = \bar{\phi}_0+  \Lambda r_g^2  \frac{3 \left(18 g_2+27 g_3-32\right)     }{g_2^3-27 g_3^2 } b(\phi)         ,
\end{equation}
where
\begin{equation}
	b(\phi) =      2 \left(   \beta(\phi) -  \beta(0) \right) ,\label{eq::b::intgr::cosmol}
\end{equation}
and $\beta(\phi)$  is defined in \ref{exact::expressions::cosmol}.
 
\subsection{Secular perturbations and comparison with other results}
For the proper time secular perturbations  we have
\begin{eqnarray}
	&  F_{0}^{C_2}= \frac{1}{M_s(4\omega_1)}\int_{P_s}  F^{C_2}(s)  d M_{s} = \frac{1}{M_s(4\omega_1)} \int_{0}^{4\omega_1}  F^{C_2}(\phi) \frac{r^2}{r_g^2} d \phi  = 0 
\end{eqnarray}
and 
\begin{eqnarray}
	F_{0}^{\bar{\phi}_0} &= \frac{1}{M_s(4\omega_1)} \int_{P_s}  F^{\bar{\phi}_0}(s)  d M_{s} = \frac{1}{M_s(4\omega_1)} 
   \int_{0}^{4\omega_1}  F^{\bar{\phi}_0}(\phi)  \frac{r^2}{r_g^2} d \phi=\nonumber \\
    &=  \Lambda r_g^2  \frac{2}{M_s(4\omega_1)} \frac{C_1}{r_g^2}  \frac{3 \left(18 g_2+27 g_3-32\right)     }{g_2^3-27 g_3^2 }
  \left( \beta(4 \omega_1) - \beta(0) \right).\label{eq::fph0::expl::cosm}
\end{eqnarray}
We do not show the explicit expression for $F_{0}^{\bar{\phi}_0} $ because it is too complicated. However, it can be easily calculated from the definition (\ref{eq::beta::lambda::app}) of $\beta(\phi)$.
Also for the coordinate  time secular perturbations
we can use the relation between the proper time and coordinate time secular corrections (\ref{eq::conect:GF}).
Due to vanishing of $F_0^{C_1}$ and $F_0^{C_2}$ there are no secular perturbations of $e$, $p$, $r_a$, $r_p$.  The only non-zero secular correction $F_{0}^{\bar{\phi}_0} $ gives an additional pericenter shift per  revolution (\ref{eq::peri::one})
\begin{equation}
	\Delta_\Lambda =   \Lambda r_g^2      \frac{3 \left(18 g_2+27 g_3-32\right)     }{g_2^3-27 g_3^2 }
  \left( \beta(4 \omega_1) - \beta(0) \right) .\label{eq::peri::one::Lambda}
\end{equation}

We compare this with the pericenter shift from the exact solution of the geodesic equations in a Schwarzschild--de Sitter space-time obtained in \cite{hackmann2008geodesic}
\begin{equation}
\Delta^{SdS} =  2\pi-2\int_{z_k}^{z_{k+1}} \frac{u du}{\sqrt{P_5(u)}},\label{eq::peri::one::HL}
\end{equation}
where we rewrite  $ P_5(u)$ in terms of our $g_2$ and $g_3$ as 
\begin{eqnarray}
	P_5(u) =& u^5-u^4   +\frac{1}{108} u^2 \left(36 \Lambda  r_g^2+9 g_2-27 g_3-4\right)\nonumber\\&+\frac{1}{108} \left(36-27 g_2\right) u^3+ \frac{\Lambda  r_g^2}{3} \frac{4 -3 g_2 }{12},
\end{eqnarray}  
and $z_i$ are the zeros of $P_5(u)$.
Since $\Lambda r_g^2$ is small and for $\Lambda=0$, $\Delta^{SdS}=\Delta^{S} =2\pi-4\omega_1$ (where $\Delta^{S}$ is pure Schwarzschild pericenter shift), we can define the part of the pericenter shift $\Delta_{\Lambda}^{HL}$ which arises when $\Lambda\neq 0$  as 
\begin{equation}
2\int_{z_k}^{z_{k+1}} \frac{x dx}{\sqrt{P_5(x)}}= 4 \omega_1 +\Delta_{\Lambda}^{HL} .\label{eq::HL::int}
\end{equation}

For the Sun--Mercury system we have  $r_g=2953.25008 ~m$, the initial conditions $r_a= 6.981708938652731\cdot 10^{10}~m$, $r_p=4.600126052898539\cdot10^{10}~m$ and taking $\Lambda=10^{-51} m^{-2}$, using numerical integration of (\ref{eq::HL::int}) we have $\Delta_{\Lambda}^{HL}=  4.04333923643109 \cdot 10^{-22}$ and from our secular correction (\ref{eq::peri::one::Lambda}) we have $\Delta_{\Lambda}=   4.0433392020640\cdot   10^{-22}  $. This demonstrates the good accuracy of our method. As we can see from Table \ref{tab:first} for initial conditions nearer to the BH horizon, our method also gives good accuracy compared to the analytical results. 
\begin{table}[h!]
    \centering
    \begin{tabular}{|c|c|c|c|c|}
    \hline
        & $\Delta_{\Lambda}    $ &  $\Delta_{\Lambda}^{HL}  $ &  $ \Delta_{\Lambda}^{PN,HL} $ &  $ \Delta_{\Lambda}^{PN,KHM} $  \\\hline
      Sun--Mercury& $ 4.04333 \cdot 10^{-22}$ &  $4.04333 \cdot   10^{-22}  $ &  $ 4.04333 \cdot 10^{-22} $ &  $   4.04333\cdot 10^{-22} $  \\\hline
      Sun--Neptune& $ 1.94306 \cdot 10^{-16}$ &  $1.94306 \cdot   10^{-16}  $ &  $ 1.94306 \cdot 10^{-16} $ &  $   1.94306\cdot 10^{-16} $  \\\hline
        $r_a=60$, $r_p=40$  & $7.9971\cdot 10^{-46} $& $8.0257  \cdot 10^{-46}$&  $ 8.0095 \cdot 10^{-46} $&  $ 7.6952 \cdot 10^{-46} $ \\\hline
         $r_a=60$, $r_p=30$ &  $ 5.64630  \cdot 10^{-46} $ &  $ 5.6686 \cdot 10^{-46} $ &  $ 5.6530 \cdot 10^{-46} $&  $ 5.3981 \cdot 10^{-46} $ \\\hline 
         $r_a=60$, $r_p=20$ &  $ 3.6855  \cdot 10^{-46} $ &  $ 3.7019 \cdot 10^{-46} $ &  $ 3.6856 \cdot 10^{-46} $&  $ 3.4824\cdot 10^{-46} $ \\\hline
         $r_a=60$, $r_p=10$ &  $ 2.0609  \cdot 10^{-46} $ &  $ 2.0714 \cdot 10^{-46} $ &  $ 2.0491 \cdot 10^{-46} $&  $  1.8853\cdot 10^{-46} $ \\\hline
         $r_a=60$, $r_p=5$ &  $ 1.3429  \cdot 10^{-46} $ &  $ 1.3504  \cdot 10^{-46} $ &  $  1.3093 \cdot 10^{-46} $&  $ 1.1495\cdot 10^{-46} $ \\\hline
    \end{tabular}
    \caption{Pericenter shifts per one revolution from our approach $\Delta_{\Lambda}$ (\ref{eq::peri::one::Lambda}), exact solution from \cite{hackmann2008geodesic}  $\Delta_{\Lambda}^{HL}$ (\ref{eq::HL::int}) and post-Newtonian results from \cite{hackmann2012observables} 
 (\ref{eq::PN::int}) and from \cite{kerr2003standard} (\ref{eq::PN::int::K})   for different initial conditions and fixed $r_g=1$, $\Lambda=10^{-51}$,  $\phi_0= 2 \omega_1 $. }
    \label{tab:first}
\end{table}

\begin{figure}[h!]
 \hspace{-3.5cm}
  \begin{subfigure}{.9\textwidth}
    \centering
    \includegraphics[width=.5\linewidth]{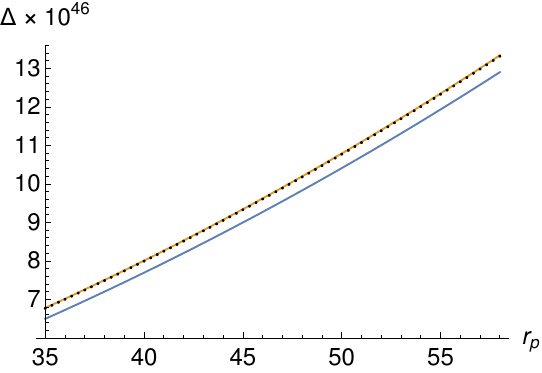}
    \caption{$r_a=60$}
    \label{fig:rp_delta_ra60_cosmol_const}
  \end{subfigure}
  \hspace{-6cm}
  \begin{subfigure}{.9\textwidth}
    \centering
    \includegraphics[width=.5\linewidth]{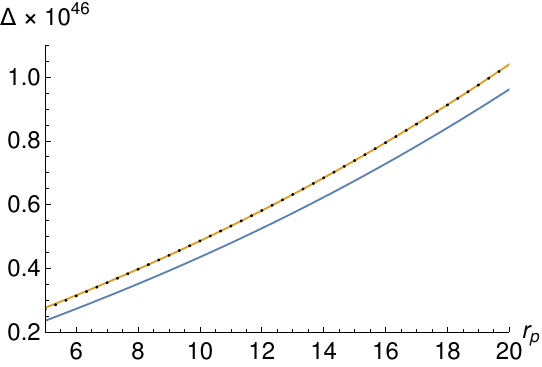}
    \caption{$r_a=30$}
    \label{fig:rp_delta_ra30_cosmol_const}
  \end{subfigure}%
  \caption{ Pericenter shifts per one revolution from our approach  $\Delta_{\Lambda}$ (\ref{eq::peri::one::Lambda}) -- black dots and from two post-Newtonian approaches: $\Delta_{\Lambda}^{HL}$ (\ref{eq::PN::int}) -- yellow line and  $\Delta_{\Lambda}^{KHM}$ (\ref{eq::PN::int::K}) -- blue line, as functions of pericenter distance $r_p$ for fixed $r_g=1$, $\Lambda=10^{-51}$, $\phi_0= 0 $ and for two different values of apocenter distance $r_a$.  }
  \label{fig:rp_delta_cosmol_const}
\end{figure}

\begin{figure}[h!]
 \hspace{-3.5cm}
  \begin{subfigure}{.9\textwidth}
    \centering
    \includegraphics[width=.5\linewidth]{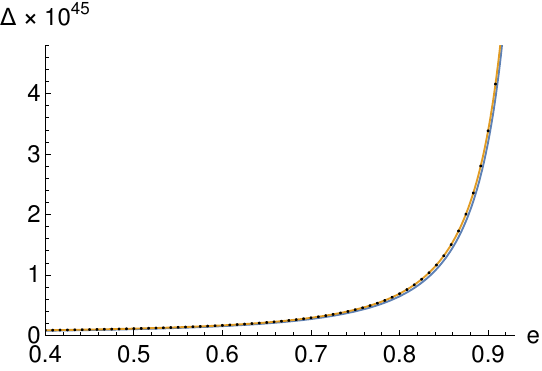}
    \caption{$p=20$}
    \label{fig:e_delta_p20_cosmol_const}
  \end{subfigure}
  \hspace{-6cm}
  \begin{subfigure}{.9\textwidth}
    \centering
    \includegraphics[width=.5\linewidth]{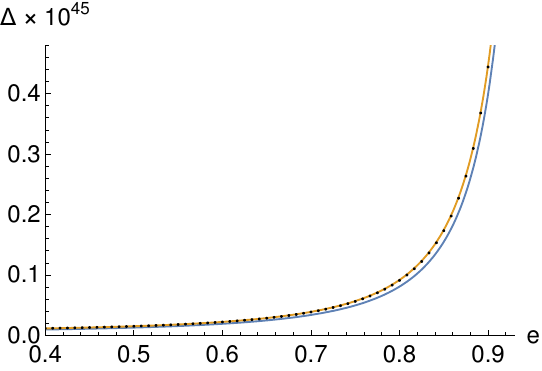}
    \caption{$p=10$}
    \label{fig:e_delta_p10_cosmol_const}
  \end{subfigure}%
  \caption{ Pericenter shifts per one revolution from our approach  $\Delta_{\Lambda}$ (\ref{eq::peri::one::Lambda}) -- black dots and from two post-Newtonian approaches: $\Delta_{\Lambda}^{HL}$ (\ref{eq::PN::int}) -- yellow line and  $\Delta_{\Lambda}^{KHM}$ (\ref{eq::PN::int::K}) -- blue line, as functions of eccentricity $e$ for fixed $r_g=1$, $\Lambda=10^{-51}$,  $\phi_0= 0 $ and for two different values of semi-latus rectum $p$.}
  \label{fig:e_delta_cosmol_const}
\end{figure}

In addition, in Table \ref{tab:first} we  compare  our result with the post-Newtonian pericenter shift per revolution obtained as an asymptotic of the exact analytical solution given in \cite{hackmann2012observables} 
\begin{equation}
\Delta_{\Lambda}^{PN,HL} =  2 \pi d^3 \sqrt{1-e^2}      \frac{    \Lambda   }{r_g} +\frac{4 \pi  \left(2-e^2\right)  d^2}{\sqrt{1-e^2}  }     \Lambda   , 
 \label{eq::PN::int}
\end{equation}
where $d=\frac{r_a+r_p}{2} $ is the semi-major axis,  
and with the result obtained in \cite{kerr2003standard} from solving the post-Newtonian analogue of Gaussian perturbation equations 
\begin{equation}
\Delta_{\Lambda}^{PN,KHM } =  2 \pi d^3 \sqrt{1-e^2}      \frac{    \Lambda   }{r_g} , 
 \label{eq::PN::int::K}
\end{equation}
which is equal to the first term in (\ref{eq::PN::int}). From Table~\ref{tab:first}, we see that for the Solar System initial data the post-Newtonian approaches work perfectly and our method agrees with them. From Table~\ref{tab:first}, Fig.~\ref{fig:rp_delta_cosmol_const} and Fig.~\ref{fig:e_delta_cosmol_const} we see that for initial data which is closer to a BH horizon our method gives at least one order advantage to post-Newtonian results obtained from solving perturbations equations (\ref{eq::PN::int::K}) and has approximately the same accuracy as the post-Newtonian asymptotic (\ref{eq::PN::int}) of the analytical solution for small eccentricity and is getting closer to the exact analytical shift (\ref{eq::HL::int}) for high eccentricity.

\begin{table}
    \centering
    \begin{tabular}{|c|c|c|c|}
    \hline
        & $\Delta_{\Lambda}$ &  $\Delta_{\Lambda}^{PN,HL}  $ &  $1- \frac{\Delta_{\Lambda}}{\Delta_{\Lambda}^{PN,HL}} $  \\\hline
      $\phi_0=0$ & $ 7.995120968946 \cdot 10^{-46}$ &  \multirow{5}{*}{ $8.009523689361 \cdot   10^{-46}  $} &  $ 0.001798199365 $ \\ \cline{1-2} \cline{4-4} 
         $\phi_0= \frac{1}{3}\omega_1$  & $7.995349480728\cdot 10^{-46} $  &  $  $&  $ 0.001769669356 $\\\cline{1-2} \cline{4-4} 
        $\phi_0= \frac{2}{3}\omega_1$   & $7.995872474588\cdot 10^{-46} $&  &  $ 0.001704372856 $\\ \cline{1-2} \cline{4-4} 
        $\phi_0= \frac{4}{3}\omega_1$   & $7.996805572986\cdot 10^{-46} $&  &  $ 0.001587874244 $\\ \cline{1-2} \cline{4-4} 
         $\phi_0=2\omega_1$  &  $ 7.997107022468 \cdot 10^{-46} $ &    &  $ 0.001550237863 $\\\hline
    \end{tabular}
    \caption{Pericenter shifts per one revolution from our approach $\Delta_{\Lambda}$ (\ref{eq::peri::one::Lambda}),  and from the post-Newtonian approach (\ref{eq::PN::int})  for different values of the argument of pericenter $\phi_0$, and $r_g=1$, $\Lambda=10^{-51}$, $r_a=60$, $r_p=40$. }
    \label{tab:second}
\end{table}

\begin{table}
    \centering
    \begin{tabular}{|c|c|c|c|}
    \hline
        & $\Delta_{\Lambda}$ &  $\Delta_{\Lambda}^{PN,HL}  $ &  $1- \frac{\Delta_{\Lambda}}{\Delta_{\Lambda}^{PN,HL}} $  \\\hline
      $\phi_0=0$ & $ 1.293073798734  \cdot 10^{-46}$ &  \multirow{5}{*}{ $1.309393097286 \cdot   10^{-46}  $} &  $ 0.012463253843 $ \\ \cline{1-2} \cline{4-4} 
         $\phi_0= \frac{1}{3}\omega_1$  & $1.328221084176\cdot 10^{-46} $  &  $  $&  $ 0.0143791707235 $\\\cline{1-2} \cline{4-4} 
        $\phi_0= \frac{2}{3}\omega_1$   & $1.340122721108 \cdot 10^{-46} $&  &  $ 0.023468600747 $\\ \cline{1-2} \cline{4-4} 
        $\phi_0= \frac{4}{3}\omega_1$   & $1.342725722866 \cdot 10^{-46} $&  &  $ 0.025456545974 $\\ \cline{1-2} \cline{4-4} 
         $\phi_0=2\omega_1$  &  $ 1.342944505857  \cdot 10^{-46} $ &    &  $ 0.0256236333005 $\\\hline
    \end{tabular}
    \caption{Pericenter shifts per one revolution from our approach $\Delta_{\Lambda}$ (\ref{eq::peri::one::Lambda}),  and from the post-Newtonian approach (\ref{eq::PN::int})  for different values of the argument of pericenter $\phi_0$, and $r_g=1$, $\Lambda=10^{-51}$, $r_a=60$, $r_p=5$. }
    \label{tab:third}
\end{table}


The pericenter shift per revolution (\ref{eq::peri::one::Lambda}) depends on $\phi_0$, in contrast to the post-Newtonian ones (\ref{eq::PN::int}) and (\ref{eq::PN::int::K}),  which is a limitation of accuracy of our method. 
In Tables \ref{tab:second} and \ref{tab:third} we compare pericenter shift per revolution from our method (\ref{eq::peri::one::Lambda}) and the post-Newtonian asymptotics of the exact solution (\ref{eq::PN::int}) for several initial values of the argument of pericenter $\phi_0$. We can see that although this dependence is quite weak for a small eccentricity (in Table \ref{tab:second} the difference is only $0.02\%$ for different $\phi_0$), it is stronger for a larger eccentricity (in Table~\ref{tab:third} the difference is $1.5\%$  for different $\phi_0$).

\section{Quantum gravitational correction to a Schwarzschild black hole}	\label{CK::21::sol}

In the effective field theory approach to quantum gravity (see \cite{donoghue2012effective})  we need to consider additional terms in Einstein--Hilbert action like 
\begin{eqnarray}
	&S = \int  d^4x \sqrt{g} \left( \Lambda+ \frac{1}{16 \pi G} R+ c_1 R^2+ c_2 R_{\mu \nu} R^{\mu \nu}  +...\right),
\end{eqnarray}
where  $R$ is the scalar curvature, $R_{\mu \nu}$ is the Ricci curvature tensor and $c_j$ are parameters of the theory which in principle can be experimentally measured. Here we neglect $\Lambda$.
As is shown in \cite{calmet2018vanishing} and \cite{calmet2017quantum} at second order in curvature,  quantum corrections do not contribute to the Schwarzschild metric. But at third order in the curvature,     
as it was obtained in \cite{CK2021}, we can write the quantum corrected Schwarzschild metric 
\begin{align}
g_{00} &=  1-\frac{r_g}{r}+ 5 \epsilon \rho^4   \frac{r_g^3}{ r^7} ,\\
g_{11} &=-\frac{1}{1-\frac{r_g}{r}+ \epsilon \rho^4   \frac{r_g^2}{r^6}\left(27 - 49 \frac{r_g}{2 r} \right) },\\
g_{22} &=-r^2,\\
g_{33} &=-r^2 \sin^2 \theta,
\end{align}
where we introduced a constant $\rho $ which has the dimension of a distance and a dimensionless parameter $\epsilon$ 
as $\epsilon \rho^4   = 128 G^2 \pi c_6$. The dimensionless parameter $c_6$ is the coefficient of the terms in the gravitational Lagrangian cubic in the curvature \cite{CK2021}. In order for the terms with $c_6$ in $g_{00}$ and $g_{11}$ to be small compared to the Schwarzschild terms, $c_6$ must be much less than $\frac{M^4 G^2}{40 \pi}$. It is convenient to define the constant $\rho$ as $\rho^4 =\frac{r_g^4}{5}$ so that $\epsilon \ll 1$ for all $c_6 \ll \frac{M^4 G^2}{40 \pi}$, which allows us to use $\epsilon$ as a small parameter for our calculations in this section. 
As in the previous section, one gets from the expansion of the  geodesic equations with respect to the small parameter $\epsilon$ the corresponding perturbation force for the Schwarzschild metric  
\begin{align}
f^\phi &=  0  ,\\
f^t &=  \epsilon  \frac{5 \rho^4  \dot{t} \dot{r} r_g^3 (7 -6 \frac{r_g}{r}) }{r^8 (1-\frac{r_g}{r})^2} \label{eq::ft::der}.
\end{align}
In analogy to the previous section, we express $f^r$ through     $f^t$ and use only  $f^t$.
In the following subsections we integrate the linearised  perturbation equations for $C_1(\phi)$, $C_2(\phi)$ and $v(\phi)$  (\ref{eqv::C1_ph}, \ref{eqv::C2_ph}, \ref{eqv::v_ph}) with this force. 

\subsection{Solutions for the osculating elements $C_1$ and $C_2$}
 After insertion of  $\dot{t}$ and $\dot{r}$ from (\ref{eqv::dot_t}) and (\ref{eqv::dot_r}) into $f^t$  (\ref{eq::ft::der})  one gets
\begin{equation}
 f^t = - \epsilon \frac{5}{2} \frac{\rho^4}{r^8} r_g^2
 \frac{C_2 C_1 \wp'(v)}{ \left(1-\frac{r_g}{r}\right)^3} \left(7-6\frac{r_g}{r}\right) .
\end{equation}
Eqns.  (\ref{eqv::C1_ph}) and (\ref{eqv::C2_ph}) for $C_1(\phi)$ and $C_2(\phi)$ with this force give
\begin{align}
C'_1(\phi) &= 0 ,\\
C'_2 (\phi) &= \left(1-\frac{r_g}{r}\right) \frac{r^2}{C_1} f^t .
\end{align}
Clearly, $C_1(\phi) = const $, so that we have only one equation 
\begin{equation}
C'_2 (\phi) =     - \epsilon \frac{5}{2} \frac{\rho^4}{r^6} r_g^2
\frac{C_2  \wp'(v)}{ \left(1-\frac{r_g}{r}\right)^2} \left(7-6\frac{r_g}{r}\right) . \label{eqv::c2_ck21}
\end{equation}
We solve (\ref{eqv::c2_ck21}) in linear approximation by $\epsilon$
	\begin{equation}
		C_2(\phi) = C_{2} + \epsilon a(\phi)  ,
	\end{equation}
which gives   
\begin{equation}
 a'(\phi) =   - \frac{5}{2} \frac{\rho^4}{r^6} r_g^2
 \frac{C_{2}  \wp'(v)}{ \left(1-\frac{r_g}{r}\right)^2} \left(7-6\frac{r_g}{r}\right)  .
\end{equation}
 After integration we have 
\begin{equation}
a(\phi) =  \frac{5  C_{2} \rho^4 }{  r_g^4  }  \left(   \alpha(\phi)  - \alpha(0) \right), 
\end{equation}
where 
\begin{eqnarray}
\alpha(\phi) =&   \wp^6(v)+ 3 \wp^5(v)+\frac{13 \wp^4(v)}{3}+\frac{113 \wp^3(v)}{27}+\frac{29 \wp^2(v)}{9 }\nonumber
\\&+\frac{181 \wp(v) }{81 }+\frac{1}{ \wp(v) -\frac{2}{3}} .
\end{eqnarray}

\subsection{Solution for for argument of pericenter }
Next we integrate the equation for the true anomaly $v$ (\ref{eqv::v_ph}) depending on $f^t$:
\begin{equation}
v'(\phi) = \frac{1 }{ 2} - \left(1-\frac{r_g}{r}\right) \frac{C_2}{C_1} \frac{2 r_g}{ \wp'(v)}  B_r^v f^t  .
\end{equation}
We rewrite this as an equation for  $\bar{\phi}_0(\phi) =\phi_0(\phi)+2\omega_3 $  instead of $v(\phi)$
\begin{equation}
\bar{\phi}_0'(\phi)=  4 \epsilon    C_2^2   \frac{5 \rho^4}{r^8}   \frac{r_g^3}{(1-\frac{r_g}{r})^2}\left(7-6\frac{r_g}{r}\right) B^v_r  , \label{eq::ph0bar}
\end{equation}
as it is more convenient to solve.
Then in linear approximation
\begin{equation}
\bar{\phi}_0(\phi) = \bar{\phi}_0(0)+ \epsilon b(\phi) ,
\end{equation}
using formulas 
(\ref{eq::int_wp_dv})-(\ref{eq::int_fracn_dv}) from  \ref{table_of_int}  we integrate (\ref{eq::ph0bar})  and obtain  
\begin{equation}
b(\phi) =   C_2^2 \frac{\rho^4}{ r_g^4} \frac{20 (3 g_2-4)}{3 \left(g_2^3-27 g_3^2\right)}     \left(\beta(\phi) -  \beta(0) \right) ,\label{eq::b::intgr}
\end{equation}
where $\beta(\phi)$  is defined in \ref{exact::expressions::QCOR}.
 
\subsection{Secular perturbations}
For  proper time secular perturbations  we have
\begin{eqnarray}
	 F_{0}^{C_2} &= \frac{1}{M_s(4\omega_1)}\int_{P_s}  F^{C_2}(s)  d M_{s} = \frac{1}{M_s(4\omega_1)} \int_{-2\omega_1}^{2\omega_1}  F^{C_2}(\phi) \frac{r^2}{r_g^2} d \phi = \nonumber\\
    &=  - \frac{1}{M_s(4\omega_1)}\int_{-2\omega_1}^{2\omega_1} \epsilon \frac{5}{2} \frac{\rho^4 }{r^6} 
 \frac{C_{2} C_{1}  \wp'(v)}{ \left(1-\frac{r_g}{r}\right)^2} \left(7-6\frac{r_g}{r}\right)  d \phi = 0 
\end{eqnarray}
and 
\begin{eqnarray}
	F_{0}^{\bar{\phi}_0} & = \frac{1}{M_s(4\omega_1)} \int_{P_s}  F^{\bar{\phi}_0}(s)  d M_{s} = \frac{1}{M_s(4\omega_1)} 
   \int_{-2\omega_1}^{2\omega_1}  F^{\bar{\phi}_0}(\phi)  \frac{r^2}{r_g^2} d \phi=\nonumber \\
    &= \epsilon  C_{2}^2 C_{1}  \frac{\rho^4}{ r_g^6}  \frac{20 (3 g_2-4)  }{ 3 \left(g_2^3-27 g_3^2\right)  } \frac{1}{M_s(4\omega_1)} 
  \left( \beta(2 \omega_1) - \beta(-2 \omega_1) \right) .\label{eq::fph0::expl}
\end{eqnarray}
We also give an explicit expression for $F_{0}^{\bar{\phi}_0} $ in \ref{exact::expressions::QCOR}.
Also for the coordinate time secular perturbations
we can use the relation between the proper time and coordinate time secular corrections (\ref{eq::conect:GF}). %

Due to vanishing of $F_0^{C_1}$ and $F_0^{C_2}$ there are no secular perturbations of $e$, $p$, $r_a$, $r_p$ so that there are no inspirals induced by quantum corrections in linear approximation.  The only non-zero secular correction $F_{0}^{\bar{\phi}_0} $ gives an additional pericenter shift per  revolution
(\ref{eq::peri::one})
that in principle may be observed for supermassive BH as in \cite{abuter2020detection}, which could allow us to estimate parameter $c_6$. As an example, for initial conditions $r_g=1$, $r_a=60$, $r_p=50$,  $\phi_0=\frac{\pi}{5}$ and $ \epsilon \rho^4 = 0.1$ we have $\Delta_{qc}= 1.176835 \cdot 10^{-7} $ and for different $r_p=10$ we have $\Delta_{qc}= 0.000227034 $.  Comparison with the pure Schwarzschild shifts $\Delta_{S}= -0.180262 $  and  $\Delta_{S}= -0.6376871 $ shows that the additional pericenter shifts are very small, even for such initial conditions when the test particle is quite close to the BH horizon. 
 
\section{Hybrid Schwarzschild/post-Newtonian $2.5$ self-force  }	\label{PP::07::solve}

There are a lot of approaches to the two-body problem in GR. One of them is the self-force technique (see \cite{barack2018self}) which can be used for the case when the two bodies have very different masses. In this approach, the smaller body moves in the metric of the larger one, but with an additional perturbation force which take into account the gravitational field of both masses. This additional so-called self-force depends on the mass ratio. In \cite{PP2007} the authors developed a version of a self-force calculation, considering a binary non-spinning system of two bodies of masses $m_1$ and $m_2$ governed by the hybrid Schwarzschild/post-Newtonian $2.5$ order equations of motion proposed in \cite{kidder1993coalescing}. Here “hybrid” means that the post-Newtonian series has two types of terms: one type depends on the mass ratio $ \frac{m_1}{m_2}$ and the other type is independent of it and equals to the first terms of the post-Newtonian expansion for the Schwarzschild metric. 
In the hybrid  approach, we take the full post-Newtonian series for the Schwarzschild metric in place of the mass ratio independent part of the post-Newtonian expansion.
In \cite{PP2007} the hybrid Schwarzschild/post-Newtonian $2.5$ equations of motion have been rewritten as a self-force problem for Schwarzschild geodesics. If the mass ratio is small, we can define a small dimensionless parameter $\epsilon= \frac{\mu}{m_1+m_2}$, where $\mu = \frac{m_1 m_2}{m_1+m_2}$ is the reduced mass. We also use $r_g = 2(m_1+m_2)$ (where we set $G=1$) and  $\mu = \frac{r_g \epsilon}{2}$. The perturbative force is given by 
\begin{align}
	f^r &=  - \frac{\mu}{r^2}\left( \mathcal{A}+\mathcal{B} \dot{r} \right) , \\
	f^\phi &=   - \frac{\mu}{r^2} \mathcal{B} \dot{\phi}  ,
\end{align}
where 
\begin{align}
	  \mathcal{A} &= \frac{\dot{t}^2}{(1-\frac{r_g}{2 r})^2} \hat{A},\\
	  \mathcal{B} &= \frac{\dot{t}^4}{(1-\frac{r_g}{2 r})^2} \left(   \frac{1}{1 -\frac{r_g}{r}} \hat{A} + \left(1-\frac{r_g}{r}\right)\hat{B}   \right),
\end{align}
and
\begin{align}
	  \hat{A} &= A_1+A_2+A_{2.5},\\
	  \hat{B} &= B_1+B_2+B_{2.5},
\end{align}
with the definitions for $A_j$ 
\begin{align}
	  A_1 &=  - 2 \frac{r_g}{2r-r_g} +3 u^2 -\frac{3}{2} \left(\frac{dr}{dt}\right)^2   ,\\
	  A_2 &= \frac{87}{4} \frac{ r_g^2}{(2r- r_g)^2} + 3 u^4 - \frac{13}{2} \frac{r_g}{2r-rg} u^2 - \frac{9}{2} u^2 \left(\frac{dr}{dt}\right)^2 \nonumber \\
      & \;\quad + \frac{15}{8} \left(\frac{dr}{dt}\right)^4 -25 \frac{r_g}{2r-r_g} \left(\frac{dr}{dt}\right)^2   ,\\
      A_{2.5} &= -\frac{8}{5} \frac{r_g}{2r-r_g}  \frac{dr}{dt} \left(\frac{17 }{3}\frac{ r_g}{ 2r- r_g}+3 u^2\right) ,
\end{align}
and for $B_j$
\begin{align}
      B_1 &= 2 \frac{dr}{dt} ,\\
	  B_2 &= - \frac{1}{2}\frac{dr}{dt} \left( 15 u^2- 41  \frac{r_g}{2r-r_g} - 9 \left(\frac{dr}{dt}\right)^2 \right),\\
      B_{2.5} &= \frac{  8  }{5}  \frac{r_g}{2r-r_g} \left( \frac{3 r_g}{2r-r_g}+u^2 \right) .
\end{align}
Here $u^2 = \delta_{ij} \frac{dx_i}{dt} \frac{dx_j}{dt} =(\frac{r_g}{2}-r)^2 \left(\frac{d \phi}{d t}\right)^2+\left(\frac{dr}{dt}\right)^2 $ is the square of the velocity in harmonic coordinates  $x_j$ which we rewrote in Schwarzchild coordinates.

\subsection{Solutions for the osculating elements $g_2$, $g_3$, $v$}
We solve perturbation equation  for $g_2$ (\ref{eqv::g2_ph}) in linear approximation in $\epsilon$ using formulas 
(\ref{eq::int_wp_dv})-(\ref{eq::int_fracn_dv}) from  \ref{table_of_int}:  
\begin{equation}
g_2(\phi) = g_{2} + \frac{\epsilon}{54} \left( 4 a_1(\phi) - \frac{C_1^2}{108 C_2^2 r_g^2}  a_2(\phi) + \frac{27 C_1}{C_2 r_g} a_{2.5}(\phi) \right),
\end{equation}
where
\begin{equation}\label{eq::alpha}
	 a_j (\phi) = \alpha_{j}( \phi) -  \alpha_{j}( 0) ,
\end{equation}
and the $\alpha_j$ are given in \ref{exact::expressions::SF}.
Analogously, for $g_3$ from  (\ref{eqv::g3_ph}) we have 
\begin{equation}
	 g_3(\phi) = g_{3} + \frac{\epsilon}{162}  \left( - 8  b_1(\phi) - \frac{C_1^2}{54  C_2^2 r_g^2} b_2(\phi)+ \frac{81 C_1}{ C_2 r_g} b_{2.5}(\phi) \right),
\end{equation}
where
\begin{equation}\label{eq::beta}
	 b_j(\phi) = \beta_{j} (\phi) - \beta_{j} (0),
\end{equation}
and the $\beta_j$ are given in \ref{exact::expressions::SF}.
Also, rewriting equation for $v$ (\ref{eqv::v_ph}) as an equation for $\bar{\phi}_0$   we have a formal solution 
\begin{equation}
\bar{\phi}_0(\phi) = \bar{\phi}_0 + \frac{\epsilon r_g}{2  }  (c_1(\phi)+c_2(\phi)+c_{2.5}(\phi)) .
\end{equation}
Owing to the presence of terms $\frac{v \wp'(v)}{1+3\wp(v)}$ and $\frac{\zeta(v) \wp'(v)}{1+3\wp(v)}$ the $c_j(\phi)$ cannot be integrated analytically but, instead, have to be calculated numerically.

\subsection{Secular perturbations}
For the proper time secular  perturbations of $g_2$ and $g_3$ we have
\begin{align}
F_{0}^{g_2} &= \frac{1}{M(4\omega_1)} \frac{\epsilon  }{2 } \frac{1}{r_g}    \frac{ C_1^2}{ C_2} a_{2.5}(4\omega_1),\\
F_{0}^{g_3} &= \frac{1}{M(4\omega_1)}   \frac{\epsilon  }{2 } \frac{1}{r_g}     \frac{C_1^2}{ C_2} b_{2.5}(4\omega_1), 
\end{align}
where due to periodicity of $a_1(\phi)$, $a_2(\phi)$ and $b_1(\phi)$, $b_2(\phi)$  there are only order $2.5$ corrections.
For $\bar{\phi}_0$ we have
\begin{equation}
F_{0}^{\bar{\phi}_0}=  \frac{1}{M(4\omega_1)} C_1 \frac{\epsilon r_g}{2  } (c_1(4\omega_1)+c_2(4\omega_1)+c_{2.5}(4\omega_1) ) .
\end{equation}
Also for coordinate  time secular perturbations,
we can use the relation between proper time and coordinate time secular corrections (\ref{eq::conect:GF}).

\begin{figure}[h!]
 \hspace{-3.5cm}
  \begin{subfigure}{.9\textwidth}
    \centering
    \includegraphics[width=.5\linewidth]{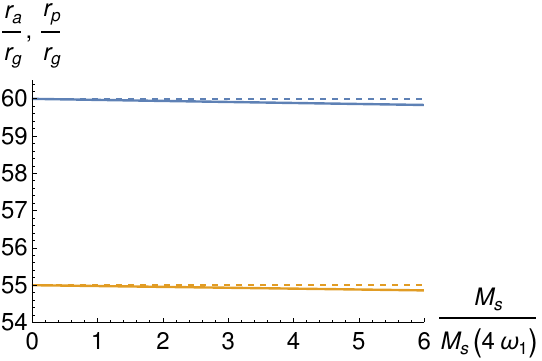}
    \caption{}
    \label{fig:secular_ra_25order_ra60_rp55_pi5}
  \end{subfigure}
  \hspace{-6cm}
  \begin{subfigure}{.9\textwidth}
    \centering
    \includegraphics[width=.5\linewidth]{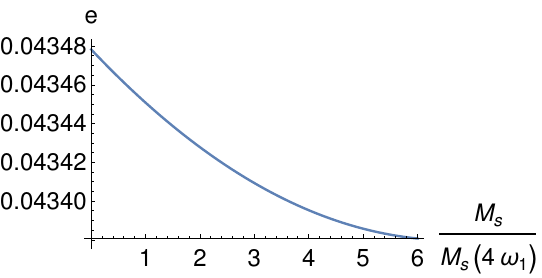}
    \caption{}
    \label{fig:secular_rp_25order_ra60_rp55_pi5}
  \end{subfigure}%
  \caption{ Secular evolution of apocenter (blue line), pericenter (yellow line) distances~(a) and eccentricity (b) with parameters $\mu = \frac{1}{10}$, $r_g=1$ and initial conditions  $\phi_0= \frac{\pi}{5} $, $r_a= 60 $, $r_p=55$.  Dashed lines are the corresponding initial conditions. The time is given in terms of the proper time mean anomaly $M_s$ in units of $M_s(4\omega_1)$.}
  \label{fig:secular_ra60_rp55_pi5}
\end{figure}

\begin{figure}[h!]
 \hspace{-3.5cm}
  \begin{subfigure}{.9\textwidth}
    \centering
    \includegraphics[width=.5\linewidth]{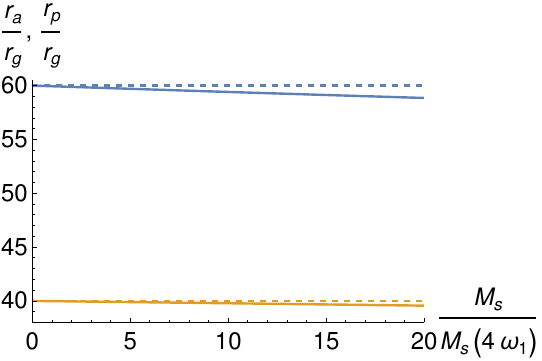}
    \caption{}
    \label{fig:secular_ra_25order_ra60_rp40_pi5}
  \end{subfigure}
  \hspace{-6cm}
  \begin{subfigure}{.9\textwidth}
    \centering
    \includegraphics[width=.5\linewidth]{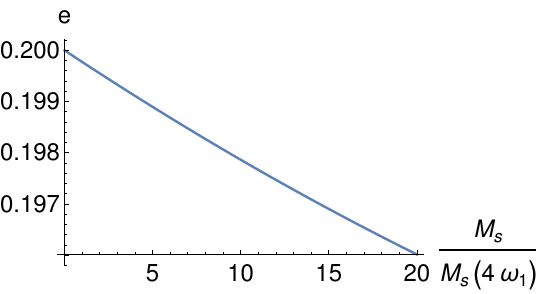}
    \caption{}
    \label{fig:secular_rp_25order_ra60_rp40_pi5}
  \end{subfigure}%
  \caption{ Secular evolution of apocenter (blue line), pericenter (yellow line) distances~(a) and eccentricity (b) with parameters $\mu = \frac{1}{10}$, $r_g=1$ and initial conditions  $\phi_0= \frac{\pi}{5} $, $r_a= 60 $, $r_p=40$.  Dashed lines are the corresponding initial conditions. The time is given in terms of the proper time mean anomaly $M_s$ in units of $M_s(4\omega_1)$.}
  \label{fig:secular_ra60_rp40_pi5}
\end{figure}

\begin{figure}[h!]
 \hspace{-3.5cm}
  \begin{subfigure}{.9\textwidth}
    \centering
    \includegraphics[width=.5\linewidth]{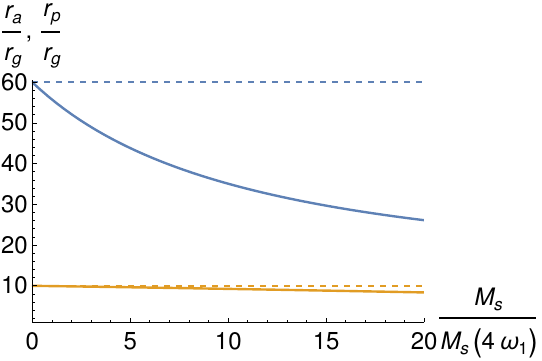}
    \caption{}
    \label{fig:secular_ra_25order_ra60_rp10_pi5}
  \end{subfigure}
  \hspace{-6cm}
  \begin{subfigure}{.9\textwidth}
    \centering
    \includegraphics[width=.5\linewidth]{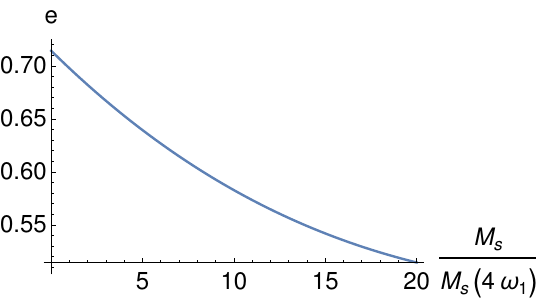}
    \caption{}
    \label{fig:secular_rp_25order_ra60_rp10_pi5}
  \end{subfigure}%
  \caption{ Secular evolution of apocenter (blue line), pericenter (yellow line) distances~(a) and eccentricity (b) with parameters $\mu = \frac{1}{10}$, $r_g=1$ and initial conditions  $\phi_0= \frac{\pi}{5} $, $r_a= 60 $, $r_p=10$.  Dashed lines are the corresponding initial conditions. The time is given in terms of the proper time mean anomaly $M_s$ in units of $M_s(4\omega_1)$.}
  \label{fig:secular_ra60_rp10_pi5}
\end{figure}

Due to non-vanishing of $F_{0}^{g_2}$ and $F_{0}^{g_3}$ we have non-zero secular corrections to $e$, $p$, $r_a$, $r_p$ which lead to inspirals as we can see in Figs. \ref{fig:secular_ra60_rp55_pi5}-\ref{fig:secular_ra60_rp10_pi5}, where we plot the secular evolution of the apocenter and pericenter distances and also the eccentricity for various initial conditions. The secular perturbation for $\bar{\phi}_0$ contributes only to an additional pericenter shift per  revolution $\Delta_{sf} $ (\ref{eq::peri::one}). Fixing the parameters and initial conditions $\mu = \frac{1}{10}$, $\phi_0= \frac{\pi}{5} $, $r_g=1$, $r_a= 60 $, $r_p=55$ we have $\Delta_{sf}= -1.7866 $, for $r_p=40$ we have $\Delta_{sf}= -0.0239023$, and for $r_p=10$ we have $\Delta_{sf}=0.0784699$. Comparison with the pure Schwarzschild shifts $\Delta_{S}= -0.170952 $,  $\Delta_{S}= -0.206085 $ and $\Delta_{S}= -0.6376871 $, respectively, one gets that these additional pericenter shifts are close or even larger than the pure shifts for such extreme initial conditions. For more realistic mass ratios the additional pericenter shifts will be much smaller.


\newpage
\section{Summary, discussion, and outlook} \label{disc::conc}

A new perturbation technique for osculating elements in a Schwarzschild space-time in terms of Weierstrass elliptic functions has been developed. The osculating elements are defined as usual. Then, restricting to the case of perturbation forces within the orbital plane, the general relativistic Gaussian perturbation equations for the osculating elements $g_2$, $g_3$, $\phi_0$ have been set up. These equations have been solved for several perturbation forces in linear approximation, leading to secular corrections of the osculating elements. The equations for the osculating elements $M_s$ and $M_t$ representing the mean proper and coordinate time anomalies are very complicated. However, in the case when the disturbing force does not depend on the proper and coordinate times explicitly, the procedure considerably simplifies and one can evaluate $M_s$ and $M_t$ directly by their definitions.

As a test model, the perturbation related to the additional influence due to the cosmological constant defined by geodesic motion in a Schwarzschild--de Sitter space-time has been considered. The linearised perturbation equations in that case have been solved. From these solutions, one obtains an additional pericenter shift. This result has been compared with two known post-Newtonian results: an asymptotics of an exact solution of the geodesic equations \cite{hackmann2012observables} and a solution of the post-Newtonian Gaussian equations \cite{kerr2003standard}. This comparison shows that the new method works better than the post-Newtonian Gaussian equations approach and gives approximately the same accuracy as the post-Newtonian approximation of the exact solution. 

As an example of possible applications of the new method, the quantum correction to the Schwarzschild metric obtained in \cite{CK2021} has been considered. This correction leads to a particular perturbation force. As another practical  application for modelling the inspiral of binary systems, the self-force within a hybrid Schwarzschild/post-Newtonian $2.5$ order formalism as introduced in \cite{PP2007} has been considered. For the corresponding perturbation force, the linearised solutions lead to secular corrections, which in principle can be observed. 

Despite the fact that in the weak field regime of, e.g., in planetary systems the post-Newtonian approach works perfectly and also is considerably simple to handle, for extreme mass-ratio inspirals and for near to BH horizon physics it is inevitable to use approaches like the new parturbation scheme presented here, or higher order post-Newtonian approximations, or the scheme presented in \cite{PP2007}. The advantage of the scheme presented in this paper is that one very quickly arrives at high precision result, however, at the cost of first doing calculations involving elliptic functions. 

Our technique could be modified for the application to the problem of motion of photons in strong gravity regime. This then would easily include the multiple orbiting of light around a BH. A further possible future direction of applications is to investigate modified GR and alternatives to GR, and their impact on geodesic motion. Also, it is interesting to consider the influence of quantum gravitational effects, non-flat asymptotics etc. on motion. One may also  consider perturbation forces not restricted to the orbital plane as they may appear for neutron stars, with additional multipole moments originating in the rotation of the stars. In order to do so, we need to expand our GR Gaussian perturbation scheme to include equations for the other osculating elements, namely the inclination $i$ and the ascending node $\Omega$. And finally, since also all geodesics around a Kerr BH are known \cite{hackmann2010geodesic} which are all also given in terms of Weierstrass elliptic functions, one may think about a perturbation theory based on the solutions of Kerr geodesics.

\section*{Acknowledgments} 

We thank Daryna Bukatova, Bennet Gr\"utzner, and Eva Hackmann for fruitful discussions. Financial support by the Deutsche Forschungsgemeinschaft (DFG, German Research Foundation) under Germany’s Excellence Strategy-EXC-2123 “QuantumFrontiers” -- Grant No. 390837967 and the CRC 1464 “Relativistic and Quantum-based Geodesy” (TerraQ) is greatfully acknowledged.


\appendix

\section{Elliptic functions}  \label{eliptic::func}
In this appendix a short introduction to elliptic functions is given which consists of the formulas that are widely used in this manuscript. For a broader view in this topic, see for example \cite{WW}, \cite{lawden2013elliptic} or other classical textbooks on elliptic functions. 

The Weierstrass elliptic $\wp$-function is a doubly periodic meromorphic function which can be defined as parametrisation of the elliptic curve  
\begin{equation}
 \wp'^2(v) = 4 \wp^3(v) - g_2 \wp(v) -g_3 = 4(\wp(v) - e_1)(\wp(v)-e_2)(\wp(v)-e_3),
\end{equation}
where $g_2$, $g_3$ are the so-called  invariants of this elliptic curve, and $e_1$, $e_2$, $e_3$ are the zeros of $\wp'$ given by $e_i= \wp(\omega_i
)$, where $\omega_i$ are  half-periods of $\wp(v)$.
It is useful to define Weierstrass zeta function $\zeta(v)$
\begin{equation}
 \wp(v)=  - \frac{d }{dv} \zeta(v) ,
\end{equation}
and Weierstrass sigma function $\sigma(v)$
\begin{equation}
 \zeta(v)=  \frac{d }{dv} \log{\sigma(v)} .
\end{equation}
 We widely use the quasi-periodic property of the $\zeta(v)$  function with quasi-periods $2\omega_i$
\begin{equation}
 \zeta(v+2\omega_i)=  \zeta(v)+ 2\eta_i ,\label{eq::shift:prop::zeta}
\end{equation}
where $\eta_i  =\zeta(\omega_i) $. 
For $\sigma(v)$ there is also a simple formula for the $2\omega_i$ shifts   
\begin{equation}
 \sigma(v+2\omega_i)=  -e^{2\eta_i(z+\omega_i)} \sigma(v).\label{eq::shift:prop::sigma}
\end{equation}

In order to use the Weierstrass $\wp$-function for physics, we need to introduce the region of $v$ where  $\wp(v)$ is a real-valued function. As is shown in \cite{lawden2013elliptic}, $\wp(v)$ is a real-valued function if and only if the half-period $\omega_1$ is real and the other half-period $\omega_3$ is imaginary. The domains where $\wp(v)$ is a real-valued function are then $v\in$  $( i x+ \omega_1)$, $(  x+ \omega_3)$, $(x)$, $( ix)$ where $x \in \mathbbm{R}$. Following  \cite{Hagihara}, in this paper the imaginary half-period shift $\wp(x+ \omega_3)$ is taken.

In addition, the derivatives of $\wp(v;g_2,g_3)$ with respect to its invariants $g_2$ and $g_3$ are used in section \ref{geodesy::equations} \cite{FunctionsWolfram}, 
\begin{eqnarray}
\frac{\partial \wp(v;g_2,g_3)}{\partial g_2} & =  \frac{2 g_2^2 \wp -36 g_3 \wp^2+\wp' \left(g_2^2 v-18 g_3 \zeta \right)+6 g_3 g_2}{4 \left(g_2^3-27 g_3^2\right)},\label{eq::part::wp::g2}
\\
\frac{\partial \wp(v;g_2,g_3)}{\partial g_3} & = \frac{12 g_2 \wp^2-18 g_3 \wp+\wp' \left(6 g_2 \zeta-9 g_3 v\right)-2 g_2^2}{2 \left(g_2^3-27 g_3^2\right)}.\label{eq::part::wp::g3}
\end{eqnarray}
Here, for simplicity, the arguments of $\wp(v;g_2,g_3)$, $\wp'(v;g_2,g_3)$, $\zeta(v;g_2,g_3)$ on the RHS have been omitted.
 
\section{Comparison of Hagihara's  and  Sharf's solutions} \label{Hagihara::Sharf}
We prove the equivalence of Hagihara's \cite{Hagihara}  and Sharf's \cite{Scharf2011}  solutions of the geodesic equations in a Schwarzschild space-time.
Let's consider Hagihara's solution 
\begin{equation}
 r = \frac{r_g}{\frac{1}{3}+  \wp(\frac{u}{2}+ \hat{\omega}_3)},
\end{equation}
where $\hat{\omega}_3$ is a half-period of $\wp(u,16  g_2, 64 g_3)$.
Using the  homogeneity of $\wp$ \cite{WW}
\begin{equation}
  \wp(\lambda u, \lambda^{-4} g_2, \lambda^{-6} g_3 ) = \lambda^{-2} \wp( u, g_2,g_3 ) ,
\end{equation}
we have 
\begin{equation}
  \wp\left(  \frac{u}{2},16  g_2, 64 g_3 \right) = 4 \wp( u, g_2,g_3 ) ,
\end{equation}
and for $r$
\begin{equation}\label{r::hag:sh}
 r = \frac{3 r_g}{1+   12 \wp(u+ \omega_3)},
\end{equation}
where $\omega_3$ is a half-period of $\wp(u,g_2, g_3)$.
In order to transform expression (\ref{r::hag:sh}) for $r$ into the corresponding expression from \cite{Scharf2011} we can use the half-period addition formula  \cite{WW}
\begin{equation}
  \wp(u+ \omega_3) = e_3 +\frac{(e_3-e_1)(e_3-e_2)}{\wp(u) -e_3},
\end{equation}
 where the zeros $e_1$, $e_2$ and $e_3$ can be expressed through  the pericenter and apocenter coordinates as defined in  \cite{Scharf2011} 
 \begin{equation}
  r_1=\frac{3 r_g}{12 e_3+1},\quad r_2=\frac{3 r_g}{12 e_2+1}.
\end{equation}
This leads to  
 \begin{equation}
  r =  r_1-\frac{3 r_1 \left(r_1-r_2\right) \left(-2 r_2 r_g-r_1 r_g+r_1 r_2\right)}{-3 r_1^2 r_g-3 r_2 r_1 r_g+3 r_2^2 r_g +3 r_2 r_1^2-2 r_2^2 r_1+12 r_2^2 r_1 \wp(u)}. 
\end{equation}
After defining $r_3 $ according to \cite{Scharf2011} 
 \begin{equation}
  r_3  = \frac{r_1 r_2 r_g}{-r_2 r_g-r_1 r_g+r_1 r_2},
\end{equation}
we finally can write  
 \begin{equation}
  r =r_1  -  \frac{3 r_1 \left(r_1-r_2\right) \left(r_1-r_3\right)}{4 \wp(u) (3 r_2 r_1  +3 r_3 r_1  +3 r_2 r_3) +3 r_1^2-2 r_2 r_1-2 r_3 r_1+r_2 r_3},
\end{equation}
which is exactly the expression from  \cite{Scharf2011}.

\section{Table of integrals} \label{table_of_int}
In this appendix, we show some integrals that appear when we solve perturbation equations with different forces.
For integrals of $\wp^n(v)$ we  use the formulas from \cite{Prudnikov:1990:IS}  
\begin{align}
    \int \wp(v) dv &=  -\zeta(v),\label{eq::int_wp_dv}\\
    \int \wp^2(v)dv &= \frac{g_2 v}{12}+\frac{\wp'(v)}{6},\label{eq::int_wp2_dv}\\
    \int \wp^3(v) dv &= -\frac{3}{20}  g_2 \zeta(v)+\frac{g_3 v}{10}+\frac{\wp(v) \wp'(v)}{10},\label{eq::int_wp3_dv}\\
    \int \wp^n(v) dv &= \frac{1}{2 (2\ n-1)}  \wp^{n-2}(v) \wp'(v) +  \frac{(2\ n-3) g_2}{4 (2\ n-1)} \int \wp^{n-2}(v)dv \nonumber \\ & \quad +  \frac{(n-2) g_3 }{2 (2\ n-1)} \int \wp^{n-3}(v)dv.\label{eq::int_wpn_dv}
\end{align}
For integrals of $  v  \wp^n(v)\wp'(v)$ we   use integration by parts  
\begin{align}
    \int  v \wp'(v) dv &=  v \wp(v) +\zeta(v) , \label{eq::int_wpP_v_dv}\\
    \int  v \wp(v) \wp'(v) dv &= \frac{ v \wp^2(v)}{2} -\frac{1}{2} \left( \frac{g_2 v}{12}+\frac{\wp'(v)}{6} \right),\label{eq::int_wpP_wp_v_dv}\\
    \int  v  \wp^n(v) \wp'(v) dv &= \frac{1}{n+1} v \wp^{n+1}(v) - \int \wp^{n+1}(v) dv \label{eq::int_wpP_wpn_v_dv} .
\end{align}
For integrals of $\zeta(v) \wp^n(v)  \wp'(v)$ we also use integration by parts  
\begin{align}
    \int  \zeta(v)  \wp'(v) dv &=  \zeta(v) \wp(v) +\int \wp^2(v)dv  ,\label{eq::int_wpP_wp_zeta_dv}\\
    \int  \zeta(v)  \wp^n(v) \wp'(v) dv &= \frac{1}{n+1}\zeta(v) \wp^{n+1}(v) + \frac{1}{n+1}\int \wp^{n+2}(v)dv.\label{eq::int_wpP_wpn_zeta_dv}
\end{align}
The same method works for integrals of the form $\wp^n(v) v $
\begin{equation}
    \int  v \wp(v) dv= - v \zeta(v) +\int \zeta(v)dv = - v \zeta(v) +\log \sigma(v) . \label{eq::int_wp_v_dv}
\end{equation}
In addition, we have integrals of $\frac{v \wp'(v)}{\left( \rho-\wp(v)\right)^n}$ and $\frac{\zeta(v) \wp'(v)}{\left(\rho-\wp(v)\right)^n}$. For $n=2 $ and $n= 3$  we get
\begin{align}
    \int \frac{v \wp'(v)}{\left(\rho-\wp(v)\right)^2} dv &=  \frac{v}{\rho-\wp(v)}+\int\frac{1}{\rho-\wp(v)} dv,\label{eq::int_wpP_v_frac2_dv}\\
    \int \frac{v \wp'(v)}{\left(\rho-\wp(v)\right)^3} dv &=  \frac{v}{2(\rho-\wp(v))^2}-\frac{1}{2 }\int\frac{1}{ (\rho-\wp(v))^2} dv,\label{eq::int_wpP_v_frac3_dv}\\
    \int  \frac{\zeta(v) \wp'(v)}{\left(\rho-\wp(v)\right)^2} dv &=  \zeta(v) \frac{1}{\rho-\wp(v)}+ \int\frac{\wp(v)}{\rho-\wp(v)} dv,\label{eq::int_wpP_zeta_frac2_dv}\\
    \int  \frac{\zeta(v) \wp'(v)}{\left(\rho-\wp(v)\right)^3} dv &=  \zeta(v) \frac{1}{2(\rho-\wp(v))^2}- \frac{1}{ 2 }\int\frac{1}{\rho-\wp(v)} dv +\frac{1}{ 2 }\int\frac{\rho}{(\rho-\wp(v))^2} dv.\label{eq::int_wpP_zeta_frac3_dv}
\end{align}
In \cite{Prudnikov:1990:IS} there are a few typos in integrals of $  \frac{1}{\left(\wp(v)-\rho \right)^n}  $. We use their correct   versions
\begin{align}
    \int \frac{1}{\wp(v)-\rho} dv &= -\frac{\xi_2(\phi,\rho)}{\sqrt{4 \rho ^3-g_2 \rho -g_3}} ,\label{eq::int_frac1_dv}\\
    \int \frac{1}{\left(\wp(v)-\rho \right)^2}  dv &= \frac{-1}{4 \rho ^3-g_2 \rho- g_3}  \left(  \xi_1(\phi,\rho)-\frac{\xi_2(\phi,\rho) \left(6 \rho ^2-\frac{g_2}{2}\right)}{\sqrt{4 \rho ^3-g_2 \rho -g_3}}  \right),\label{eq::int_frac2_dv}\\
    \int \frac{1}{\left(\wp(v)-\rho \right)^n}   dv &= \frac{\wp'(v)}{(\wp(v)-\rho )^{n-1} }\frac{1}{\left((n-1) \left( g_2 \rho +g_3-4 \rho ^3\right)\right)}-\nonumber \\
    & \quad -\frac{\left((2 n-3) \left(g_2-12 \rho ^2\right)\right)} {2 (n-1) \left(g_2 \rho +g_3-4 \rho ^3\right)} \int  \frac{1}{\left(\wp(v)-\rho \right)^{n-1}} dv \nonumber\\
    &\quad + \frac{(12 (n-2) \rho ) }{(n-1) \left(g_2 \rho +g_3-4 \rho ^3\right)}\int  \frac{1}{\left(\wp(v)-\rho \right)^{n-2}} dv \nonumber\\
    &\quad -\frac{(2 (5-2 n)) }{(n-1) \left(g_2 \rho +g_3-4 \rho ^3\right)} \int  \frac{1}{\left(\wp(v)-\rho \right)^{n-3}} dv .\label{eq::int_fracn_dv}
\end{align}

\section{Exact expressions} \label{exact::expressions} 
In this part of the appendix, we present some exact expressions and definitions that we used in the main sections.

\subsection{Exact expressions for $M_s(4\omega_1)$ and $M_t(4\omega_1)$ } \label{exact::expressions::MsMt} 
For proper time mean anomaly, using the $2\omega_1$ shift properties (\ref{eq::shift:prop::zeta}), (\ref{eq::shift:prop::sigma})  of $\sigma(v)$ and~$\zeta(v)$ functions  we can calculate  
\begin{equation}
M_s(4\omega_1) =  \frac{-2}{\wp'(y)}\left( 4 \eta _1-\frac{3 \wp''(y) (4y \eta _1+2 i \pi)+4 \omega _1 (\wp'(y)+3 \wp''(y) \zeta(y))}{3 \wp'(y)} \right)
\end{equation}
and also for coordinate time mean anomaly 
\begin{equation}
M_t(4\omega_1) = M_s(4\omega_1)+   8\frac{\omega_1 \zeta(y)-  \eta_1 y }{ \wp'(y)}-  8\frac{\omega_1 \zeta(z)-  \eta_1 z }{ \wp'(z)}.
\end{equation}

\subsection{Exact expression for $\beta(\phi)$ for the Schwarzschild--de Sitter metric} \label{exact::expressions::cosmol} 
After integration of (\ref{eq::ph0bar::cosmol}) and simplification,  we obtain for the $\beta$ introduced in (\ref{eq::b::intgr::cosmol})  
\begin{eqnarray}
 \beta(\phi) = &\beta_{v} v + \beta_{\zeta} \zeta + \beta_{\wp}  \wp' + \beta^1_1 \xi_1(\phi,-1/3)+\nonumber\\&+\beta^1_2 \xi_2(\phi,-1/3)+\beta^2_1 \xi_1(\phi,2/3)+\beta^2_2 \xi_2(\phi,2/3),\label{eq::beta::lambda::app}
\end{eqnarray}
where 
\begin{align}
\beta_{v} &= \frac{g_3}{-27 \wp ^3+9 \wp +2}+\frac{4 \left(3 g_2 \left(3 g_2-5\right)+4\right)}{81 \left(9 g_2-27 g_3-4\right)}+\frac{4}{81}, \\
\beta_{\zeta} &= -\frac{2 g_2}{-81 \wp ^3+27 \wp +6}, \\
\beta_{\wp} &= \frac{2 \left(g_2 \left(3 g_2-2\right)-9 g_3\right)}{3 \left(9 g_2-27 g_3-4\right) (3 \wp +1)^2}, \\      
\beta^1_1 &= \frac{297 g_2^3-6 \left(27 g_3+28\right) g_2^2+\left(64-702 g_3\right) g_2+9 g_3 \left(56-351 g_3\right)}{27 \left(-9 g_2+27 g_3+4\right){}^2},\\ 
\beta^1_2 &= \frac{\left(99 g_2-54 g_3-116\right) \left(g_2^3-27 g_3^2\right)}{2 \sqrt{3} \left(9 g_2-27 g_3-4\right){}^{5/2}},\\
\beta^2_1 &= \frac{2 \left(g_2 \left(3 g_2-8\right)+18 g_3\right)}{27 \left(18 g_2+27 g_3-32\right)},\\ 
\beta^2_2 &= -\frac{g_2^3-27 g_3^2}{\sqrt{3} \left(-18 g_2-27 g_3+32\right){}^{3/2}},
\end{align}
where we omitted the argument of $\wp(v)$, $\wp'(v)$ and $\zeta(v)$  for simplicity.

\subsection{Exact expression for $\beta(\phi)$ for the quantum correction} \label{exact::expressions::QCOR} 
After integration of (\ref{eq::ph0bar}) and simplification  we obtain $\beta$ introduced in (\ref{eq::b::intgr})  as  
\begin{equation}
 \beta(\phi) = \beta_{v} v + \beta_{\zeta} \zeta + \beta_{\wp}  \wp' + \beta_1 \xi_1(\phi,3/2)+\beta_2 \xi_2(\phi,3/2),
\end{equation}
where   
\begin{align}
  \beta_{v} &=  \frac{9  }{\frac{2}{3}-\wp } -\frac{1}{9} \wp  (3 \wp  (\wp  (9 \wp  (3 \wp  (\wp +3)+13)+113)+87)+181) \nonumber\\ 
 & \quad +\frac{1}{960} g_2 \left(69 g_2^2+260 g_2+2320\right)+\frac{1}{2592 } \frac{g_2}{  g_3} \left(g_2 \left(3 g_2 \left(135 g_2+452\right)-2896\right)-15552\right)\nonumber\\ 
 & \quad -\frac{9}{5}  g_3^2-\frac{1}{6}  g_3 \left(27 g_2+113\right), \\
 \beta_{\zeta} &= -\frac{6  }{\frac{2}{3}-\wp } +\frac{2}{27} \wp  (3 \wp  (\wp  (9 \wp  (3 \wp  (\wp +3)+13)+113)+87)+181)  +\left(\frac{3 g_3}{16}-\frac{29}{18}\right) g_2 \nonumber
 \\ & \quad +\frac{g_3 \left(351 g_3+181\right)}{9 g_2}+\frac{1}{180} g_3 \left(621 g_3+565\right) -\frac{21}{160}  g_2^3-\frac{13 g_2^2}{8},
 \end{align}
 \begin{align}
 \beta_{\wp} &= 3 g_2 \wp ^5+ 9 \left( g_2-\frac{ g_3}{2}\right) \wp ^4+\left(\frac{ g_2^2}{4}+13 g_2-\frac{27 g_3}{2}\right) \wp ^3 \nonumber
 \\ & \quad +\left(\frac{3 g_2^2}{4}-\frac{3 g_3 g_2}{8}+\frac{113 g_2}{9}-\frac{39 g_3}{2}\right) \wp^2 \nonumber
 \\& \quad +\left(\frac{7 g_2^3}{80}+\frac{13 g_2^2}{12}+\frac{29 g_2}{3}-\frac{9 g_3 g_2}{8}-\frac{9}{5} g_3^2-\frac{113 g_3}{6}\right) \wp \nonumber
 \\& \quad +\frac{5 g_2^3}{16}+\frac{113 g_2^2}{108}+\frac{181 g_2}{27}-\frac{1}{32} g_2^2 g_3-\frac{29 g_3}{2}-\frac{27 g_3^2}{4}-\frac{13 g_2 g_3}{8} ,\\       
  \beta_1 &= \frac{18 \left(3 g_2^2-8 g_2+18 g_3\right)}{18 g_2+27 g_3-32},\\ 
  \beta_2 &= \frac{27  \left(g_2^3-27 g_3^2\right)}{\left(18 g_2+27 g_3-32\right) \wp'(y)} ,
\end{align}
where we omitted the argument of $\wp(v)$, $\wp'(v)$ and $\zeta(v)$  for simplicity.

For the proper time secular perturbation  (\ref{eq::fph0::expl}) we have an explicit expression 
\begin{align}
F_{0}^{\bar{\phi}_0}  &= \epsilon  C_{2}^2 C_{1}  \frac{\rho^4}{ r_g^6}  \frac{20 (3 g_2-4)  }{ 3 \left(g_2^3-27 g_3^2\right)  } \frac{1}{M_s(4\omega_1)} 
 \nonumber 
  \\ & \quad \times \left(    \eta _1 \left(12 \hat{e}_2^6 g_2+36 \hat{e}_2^5 g_2+52 \hat{e}_2^4 g_2+\frac{452}{9} \hat{e}_2^3 g_2+\frac{116}{3} \hat{e}_2^2 g_2+\frac{724 \hat{e}_2 g_2}{27} \right.\right. \nonumber 
  \\ & \quad \left.\left.+36 g_2 \left(\frac{1}{3 \hat{e}_2-2}-\frac{16}{18 g_2+27 g_3-32}\right)+\frac{1}{90} g_2 g_3 \left(621 g_3+565\right) -\frac{13 g_2^3}{4}-\frac{21 g_2^4}{80}\right.\right. \nonumber 
  \\ & \quad \left.\left.+g_2^2 \left(\frac{3 g_3}{8}+\frac{216}{18 g_2+27 g_3-32}-\frac{29}{9}\right)+\frac{2}{9} g_3 \left(351 g_3+\frac{5832}{18 g_2+27 g_3-32}+181\right) \right) \right. \nonumber 
  \\ & \quad \left. +\omega _1 \left(-\frac{2}{9} \hat{e}_2 \left(3 \hat{e}_2 \left(\hat{e}_2 \left(9 \hat{e}_2 \left(3 \hat{e}_2 \left(\hat{e}_2+3\right)+13\right)+113\right)+87\right)+181\right) g_3 \right.\right. \nonumber 
  \\ & \quad \left.\left.+\frac{54 g_3}{2-3 \hat{e}_2}-\frac{108 \zeta(y) \left(g_2^3-27 g_3^2\right)}{\left(18 g_2+27 g_3-32\right) \wp '\left(y\right)}+\frac{5 g_2^4}{16}+\frac{113 g_2^3}{108}-\frac{181 g_2^2}{81} \right.\right. \nonumber 
  \\ & \quad \left.\left.+\frac{1}{480} \left(g_2 \left(69 g_2+260\right)+2320\right) g_3 g_2-12 g_2-\frac{1}{3} \left(27 g_2+113\right) g_3^2-\frac{18 g_3^3}{5} \right.\right. \nonumber 
  \\ & \quad \left.\left.+\frac{16 \left(9 g_2^2-60 g_2+64\right)}{18 g_2+27 g_3-32}+32\right) +\frac{108 \left(g_2^3-27 g_3^2\right) \wp ^{-1}\left(\frac{2}{3}\right) \eta _1}{\left(18 g_2+27 g_3-32\right) \wp '\left(y\right)}   \right) ,
\end{align}
where $\hat{e}_2=\wp(\frac{\phi_0}{2} +\omega_2)$.

\subsection{Exact expressions for $\alpha_j (v)$ and $ \beta_j (v) $ for the self-force} \label{exact::expressions::SF} 
For the functions $\alpha_j$ defined in (\ref{eq::alpha}) we have: for $\alpha_1$ 
 \begin{align}
\alpha_1 (v) &= \frac{16 \left(18 g_2+27 g_3-32\right)}{ \wp -2/3}+\frac{2 \left(18 g_2+27 g_3-32\right)}{( \wp -2/3)^2}+\frac{ \left(153 g_2+351 g_3+988\right)}{ \wp -5/3} \nonumber \\
& \quad -\frac{4 \left(18 g_2+27 g_3-32\right)}{( \wp -5/3)^2}+1125 -81 \wp ^2-540 \wp \\ 
& \quad + 261 \left(2 g_2+3 g_3\right) \log \frac{2/3- \wp}{5/3- \wp} -766 \log (2-3 \wp )+2224 \log (5-3 \wp ),\nonumber 
\end{align}
for $\alpha_2$      
\begin{align}
\alpha_2(v) &= \frac{12}{5} \left(3 g_2-964\right) (5-3 \wp )^5-4 \left(36 g_2-27 g_3-4441\right) (5-3 \wp )^4 \nonumber \\
& \quad + 4 \left(261 g_2-702 g_3-26168\right) (5-3 \wp )^3 \nonumber \\
& \quad + \frac{3}{2} \left(9 g_2 \left(9 g_2-296\right)+4 \left(5157 g_3+82268\right)\right) (5-3 \wp )^2 \nonumber \\
& \quad + 9 \left(27 g_2 \left(3 g_2+18 g_3+728\right)+4 \left(783 g_3-66484\right)\right) (5-3 \wp )\nonumber \\
& \quad + \frac{36 \left(18 g_2+27 g_3-32\right) \left(1035 g_2+4104 g_3+24620\right)}{(5-3 \wp )^2}  \nonumber \\
& \quad +\frac{3 \left(813483 g_2^2+18 \left(125739 g_3-480476\right) g_2+27 g_3 \left(58131 g_3-465032\right)+13333424\right)}{5-3 \wp} \nonumber \\
& \quad +\frac{6264 \left(18 g_2+27 g_3-32\right){}^2}{(5-3 \wp )^3}+\frac{3132 \left(18 g_2+27 g_3-32\right){}^2}{2-3 \wp}  -\frac{783 \left(18 g_2+27 g_3-32\right){}^2}{(2-3 \wp)^2} \nonumber \\
& \quad -2 \left(18 g_2+27 g_3-32\right) \left(28962 g_2+43443 g_3-50786\right) \log (2-3 \wp) \nonumber \\
& \quad -\left(1060371 g_2^2+1656 \left(1917 g_3-463\right) g_2+27 g_3 \left(87453 g_3-57392\right)-4664464\right) \log (5-3 \wp ) \nonumber \\
& \quad + \frac{1}{3} (5-3 \wp)^8+216 (5-3 \wp )^6-\frac{88}{7} (5-3 \wp)^7  ,
\end{align}
for $\alpha_{2,5}$    
\begin{align}
\alpha_{2.5}(v) &= \alpha_{2.5}^v  v + \alpha_{2.5}^\zeta \zeta + \alpha_{2.5}^\wp \wp'+ \alpha_{2.5}^1 \xi_1(v,2/3 )    +\alpha_{2.5}^2 \xi_2(v, 2/3) \nonumber  \\ 
& \quad + \alpha_{2.5}^{1,5} \xi_1(v, 5/3 )    +\alpha_{2.5}^{2, 5} \xi_2(v,5/3 ) ,
\end{align}
where
\begin{align}
\alpha_{2.5}^v  &=  \frac{6912 \left(3 g_2-100\right) \left(52-3 g_2\right){}^2}{\left(45 g_2+27 g_3-500\right){}^2}+\frac{64}{225} \left(3 g_2+2680\right) g_3  \nonumber \\
& \quad -\frac{32 \left(3 g_2 \left(1791 g_2-392392\right)+23475536\right)}{8505}  \nonumber\\
& \quad -\frac{256 \left(3 g_2-52\right) \left(159 g_2-7892\right)}{15 \left(45 g_2+27 g_3-500\right)} , \\ 
\alpha_{2.5}^\zeta &=  -\frac{32 \left(81 g_2^2+73620 g_2-2139200\right)}{2025}-\frac{18688 g_3}{105}, 
 \end{align}
 \begin{align}
\alpha_{2.5}^\wp &=   \frac{32}{75} \left(9 g_2-940\right) \wp -\frac{224 g_2 \wp }{75}+   \frac{512 \left(387 g_2-4468\right)}{45 (3 \wp -5)^3}-\frac{1024 \left(3 g_2-100\right)}{3 (3 \wp -5)^2} \nonumber  \\
& \quad  -\frac{64 \left(63 g_2+38924\right)}{45 (3 \wp -5)^2}   -\frac{384 g_3 (57 \wp -127)}{5 (3 \wp -5)^3} \nonumber\\ 
& \quad + \frac{31104 \left(27 g_2^3-1836 g_2^2+39312 g_2-270400\right)}{\left(45 g_2+27 g_3-500\right){}^2 (3 \wp -5)^2} \nonumber\\
& \quad -\frac{384 \left(1431 g_2^2 \wp -95832 g_2 \wp -1413 g_2^2+126024 g_2+1231152 \wp -1759888\right)}{5 \left(45 g_2+27 g_3-500\right) (3 \wp -5)^3} \nonumber\\
& \quad +\frac{11456 g_2}{315}-\frac{128 \wp ^3}{15}-\frac{512 \wp ^2}{7}-\frac{168448}{81},\\
\alpha_{2.5}^{1,2} &= \frac{2048}{405} \left(9 g_2-16\right)+\frac{1024 g_3}{15}, \\
 \quad \alpha_{2.5}^{2,2} &= \frac{4096}{405} \left(9 g_2+116\right)+\frac{4736 g_3}{15}-\frac{96 \left(69 g_2^2-17768 g_2+382480\right)}{5 \left(45 g_2+27 g_3-500\right)}  \nonumber\\
& \quad  +\frac{576 \left(1431 g_2^3-169452 g_2^2+5324112 g_2-48825920\right)}{5 \left(45 g_2+27 g_3-500\right){}^2} \nonumber\\
& \quad -\frac{46656 \left(81 g_2^4-8208 g_2^3+301536 g_2^2-4742400 g_2+27040000\right)}{\left(45 g_2+27 g_3-500\right){}^3},\\
\alpha_{2.5}^{1,5} &=  \frac{128 g_2 \sqrt{-18 g_2-27 g_3+32}}{\sqrt{3}}+\frac{2048 g_3 \sqrt{-18 g_2-27 g_3+32}}{15 \sqrt{3}}
\nonumber\\
& \quad -\frac{145408 \sqrt{-18 g_2-27 g_3+32}}{405 \sqrt{3}} , \\
\alpha_{2.5}^{2,5} &= -\frac{16}{405 \sqrt{3} \left(-45 g_2-27 g_3+500\right)^{7/2}} \Bigl(10914912405 g_2^5 \nonumber\\ 
& \quad +26244 \left(1468746 g_3-24867475\right) g_2^4\nonumber\\
& \quad +729 \left(74323737 g_3^2-2426274432 g_3+22294720480\right) g_2^3 \nonumber\\
& \quad +78732 \left(484056 g_3^3-22947651 g_3^2+404829184 g_3-2597071200\right) g_2^2 \nonumber\\
& \quad +72 \left(184410027 g_3^4-11573328438 g_3^3+294821601174 g_3^2-3524912553600 g_3\right. \nonumber\\
& \quad\qquad \left.+18327095780000\right) g_2  \nonumber\\
& \quad  +32 \left(57395628 g_3^5 - 4397674275 g_3^4 + 158978764314 g_3^3 \right.\nonumber\\
& \quad\qquad \left.-2403878449158 g_3^2+25048687752000 g_3-107232297500000\right)\Bigr),
\end{align}
where we omitted the argument of $\wp(v)$, $\wp'(v)$ and $\zeta(v)$  for simplicity.

Regarding the functions $\beta_j$ defined in (\ref{eq::beta}) we have for $\beta_1$
\begin{align}
\beta_1(v) &= \frac{16 \left(18 g_2+27 g_3-32\right)}{ \wp -2/3}+\frac{2 \left(18 g_2+27 g_3-32\right)}{( \wp -2/3)^2} \nonumber \\ 
& \quad +\frac{ \left(45 g_2+189 g_3+1180\right)}{ \wp -5/3}-\frac{4 \left(18 g_2+27 g_3-32\right)}{( \wp -5/3)^2}  -81 \wp ^2-540 \wp +1125 \nonumber\\ 
& \quad + 261 \left(2 g_2+3 g_3\right) \log \frac{2/3- \wp}{5/3- \wp }-766 \log (2-3 \wp )+2224 \log (5-3 \wp ) , 
\end{align}
and for $\beta_2$     
\begin{align}
\beta_2(v) &= \frac{2}{5} \left(72 g_2+135 g_3+5624\right) (5-3 \wp )^5-\left(882 g_2+1647 g_3 + 15940\right) (5-3 \wp )^4 \nonumber\\
& \quad + 2 \left(6012 g_2+11205 g_3+40720\right) (5-3 \wp )^3 \nonumber\\ 
& \quad + \frac{3}{2} \left(81 g_2 \left(7 g_2+12 \left(g_3-70\right)\right) - 4 \left(31239 g_3+51356\right)\right) (5-3 \wp )^2 \nonumber\\ 
& \quad +\frac{6264 \left(18 g_2+27 g_3-32\right)^2}{(3 \wp -5)^3}+\frac{3132 \left(18 g_2+27 g_3-32\right)^2}{3 \wp -2}+\frac{783 \left(18 g_2+27 g_3-32\right){}^2}{(2-3 \wp )^2} \nonumber\\ 
& \quad + 3 \left(7371 g_2^2+18 \left(783 g_3-9868\right) g_2+108 g_3 \left(27 g_3-3055\right)-399728\right) (3 \wp -5) \nonumber\\ 
& \quad + \frac{2113857 g_2^2+54 \left(109053 g_3-443852\right) g_2+81 g_3 \left(50517 g_3-430136\right)+37516560}{3 \wp -5} \nonumber\\ 
& \quad -\frac{126 \left(18 g_2+27 g_3-32\right) \left(612 g_2+1647 g_3+6472\right)}{(5-3 \wp )^2} \nonumber\\
& \quad +2 \left(18 g_2+27 g_3-32\right) \left(28962 g_2+43443 g_3-50786\right) \log (2-3 \wp ) \nonumber\\
& \quad + \left(1136187 g_2^2+36 \left(92313 g_3-59242\right) g_2+27 g_3 \left(89397 g_3 - 128096\right)-2475664\right) \nonumber\\ 
& \quad \times \log (5-3 \wp )  -\frac{1}{3} (5-3 \wp )^8+\frac{88}{7} (5-3 \wp )^7-216 (5-3 \wp )^6   ,
\end{align}
and finally for $\beta_{2.5}$
\begin{align}
\beta_{2.5}(v) &= \beta_{2.5}^v  v + \beta_{2.5}^\zeta \zeta + \beta_{2.5}^\wp \wp'+ \beta_{2.5}^1 \xi_1(v, 2/3)    +\beta_{2.5}^2 \xi_2(v, 2/3)\nonumber\\
& \quad +\beta_{2.5}^{1,5} \xi_1(v, 5/3 )    +\beta_{2.5}^{2,5} \xi_2(v,5/3 ) ,
\end{align}
with the coefficients 
\begin{align}
\beta_{2.5}^v  &= \frac{32}{42525 \left(45 g_2+27 g_3-500\right)^2} \nonumber\\ 
& \quad \times  \left(24786000 g_2^4+13286025 g_3 g_2^3-7005301200 g_2^3-7381125 g_3^2 g_2^2-7143529320 g_3 g_2^2 \right.\nonumber\\ 
& \quad \left.+293481986400 g_2^2-1791153 g_3^3 g_2-1532737080 g_3^2 g_2+253449861840 g_3 g_2 \right.\nonumber\\
& \quad \left.-4001472294400 g_2 + 1240029 g_3^4 + 228165336 g_3^3 + 54675991440 g_3^2 \right.\nonumber\\
& \quad \left.-1837155075840 g_3+17815770560000\right), \\ 
\beta_{2.5}^\zeta & =  -\frac{16 \left(1134 g_2^2+63 \left(81 g_3-25016\right) g_2+421740 g_3+59024000\right)}{42525} ,
\end{align}
\begin{align}
\beta_{2.5}^\wp &=  \frac{32 g_3 \left(27 \wp ^4+135 \wp ^3-1125 \wp ^2+4135 \wp -3720\right)}{25 (3 \wp -5)^3}+\frac{64}{42525 (3 \wp -5)^3} \nonumber\\
& \quad \times \left(5103 g_2 \wp ^4-307395 g_2 \wp ^3+1451925 g_2 \wp ^2-3342195 g_2 \wp +358110 g_2+102060 \wp ^6 \right.\nonumber\\ 
& \quad \left.+364500 \wp ^5+1246104 \wp ^4+7569720 \wp ^3-87304500 \wp ^2+287944020 \wp -236745040\right)\nonumber\\
& \quad +\frac{512 \left(2538 g_2^2 \wp -140256 g_2 \wp -3015 g_2^2+191640 g_2+1668576 \wp -2415920\right)}{5 \left(45 g_2+27 g_3-500\right) (3 \wp -5)^3} \nonumber\\ 
& \quad - \frac{51840 \left(27 g_2^3-1836 g_2^2+39312 g_2-270400\right)}{\left(45 g_2+27 g_3-500\right){}^2 (3 \wp -5)^2} \\
\beta_{2.5}^{1,2} &= -\frac{2048 \left(18 g_2+27 g_3-32\right)}{1215}, \\
\beta_{2.5}^{1,5} & = -\frac{32}{1215 \left(45 g_2+27 g_3-500\right){}^3}   \left(196961220 g_2^4+729 \left(1323243 g_3+678040\right) g_2^3 \right. \nonumber\\ 
& \quad \left.+486 \left(2641167 g_3^2-35923554 g_3-446818960\right) g_2^2\right. \nonumber\\ 
& \quad \left.+9 \left(81271107 g_3^3-2499601032 g_3^2-16343517744 g_3+293768080000\right) g_2\right. \nonumber\\ 
& \quad \left.+2 \left(64304361 g_3^4-5370900210 g_3^3-44464404384 g_3^2 \right.\right. \nonumber\\
& \quad \left.\left.+393527656800 g_3-5409999200000\right)\right), \\
\beta_{2.5}^{1,2} &= \frac{128 \left(11664 g_2^2+18 \left(1593 g_3-3136\right) g_2+16767 g_3^2-73440 g_3+63488\right)}{1215 \sqrt{-54 g_2-81 g_3+96}} ,\\
\beta_{2.5}^{2,5} &= \frac{16}{1215 \sqrt{3} \left(-45 g_2-27 g_3+500\right){}^{7/2}} \Bigl(16889194980 g_2^5 \nonumber\\ 
& \quad + 6561 \left(8629443 g_3-161903960\right) g_2^4 \nonumber\\
& \quad +1458 \left(52545591 g_3^2-1844441712 g_3+19415365120\right) g_2^3 \nonumber\\ 
& \quad +243 \left(217280637 g_3^3-10655483904 g_3^2+215931305952 g_3-1522357408000\right) g_2^2 \nonumber\\ 
& \quad +18 \left(998577639 g_3^4-63667497852 g_3^3+1892672945136 g_3^2 \right. \nonumber\\
& \quad\qquad \left.-23511911558400 g_3+139034680000000\right) g_2 \nonumber \\
& \quad + 4 \left(617003001 g_3^5-42433438086 g_3^4+2238153693804 g_3^3 \right. \nonumber\\
& \quad\qquad  \left.-27834792540480 g_3^2+358914076920000 g_3-1680572600000000\right) \Bigr),
\end{align}
where we omitted the argument of $\wp(v)$, $\wp'(v)$ and $\zeta(v)$  for simplicity.

\section*{References}
\bibliography{bib_GR}
\end{document}